\documentclass[aps,pra,superscriptaddress, nofootinbib,reprint]{revtex4-1}

\usepackage{graphicx}
\usepackage{makecell}
\usepackage{lineno}
\usepackage{amsmath}
\usepackage[table]{xcolor}
\usepackage{ulem}
\usepackage{physics}

\newcommand{\change}[1]{{\color{black}#1}}

\begin{document}

\title{Quantum computation in a hybrid array of molecules and Rydberg atoms}

\author{Chi Zhang}
\email[]{c.zhang@imperial.ac.uk}
\altaffiliation[Present address: ]{Division of Physics, Mathematics, and Astronomy, California Institute of Technology, Pasadena, CA 91125, USA. chizhang@caltech.edu}
\affiliation{Centre for Cold Matter, Blackett Laboratory, Imperial College London, Prince Consort Road, London SW7 2AZ, United Kingdom}

\author{M. R. Tarbutt}
\email[]{m.tarbutt@imperial.ac.uk}
\affiliation{Centre for Cold Matter, Blackett Laboratory, Imperial College London, Prince Consort Road, London SW7 2AZ, United Kingdom}

\begin{abstract}

We show that an array of polar molecules interacting with Rydberg atoms is a promising hybrid system for scalable quantum computation. Quantum information is stored in long-lived hyperfine or rotational states of molecules which interact indirectly through resonant dipole-dipole interactions with Rydberg atoms. A two-qubit gate based on this interaction has a duration of 1~$\mu$s and an achievable fidelity of 99.9\%. The gate \change{has little sensitivity} to the motional states of the particles -- the molecules can be in thermal states, the atoms do not need to be trapped during Rydberg excitation, the gate does not heat the molecules, and heating of the atoms \change{has a negligible effect}. Within a large, static array, the gate can be applied to arbitrary pairs of molecules separated by tens of micrometres, making the scheme highly scalable. The molecule-atom interaction can also be used for rapid qubit initialization and efficient, non-destructive qubit readout, without driving any molecular transitions. Single qubit gates are driven using microwave pulses alone, exploiting the strong electric dipole transitions between rotational states. Thus, all operations required for large scale quantum computation can be done without moving the molecules or exciting them out of their ground electronic states.

\end{abstract}

\maketitle

\section{Introduction}

Ultracold molecules are an emerging new tool for quantum simulation and information processing. They can be produced by direct laser cooling \cite{Barry2014, Truppe2017b, Cheuk2018, Caldwell2019, Ding2020, Langin2021}, by assembly of ultracold atoms \cite{Ni2008, DeMarco2019}, or by optoelectrical Sisyphus cooling~\cite{Prehn2016}. The rotational and spin degrees of freedom provide a large set of stable states to form qubits or qudits~\cite{Sawant2020} that are easily manipulated using microwave pulses. This high-dimensional space could be used to encode error-corrected qubits~\cite{Albert2020}. Coherence times of several seconds have been demonstrated for hyperfine states~\cite{Park2017,Gregory2021} and hundreds of milliseconds for rotational states~\cite{Burchesky2021} of molecules.  Single molecules have been trapped in optical tweezers \cite{Anderegg2019, Liu2019, Zhang2020, Cairncross2021}, and a tweezer array of molecules is a particularly attractive platform for quantum computing. Single-site addressing and readout is straightforward in these arrays, and the array can be reconfigured to remove defects and target interactions between chosen pairs. Dipole-dipole interactions can be used to implement two-qubit gates between molecules, and many gate protocols have been suggested~\cite{DeMille2002, Yelin2006, Pellegrini2011, Wei2016, Ni2018, Hudson2018, Hughes2020, Caldwell2020c}. However, these gate schemes are sensitive to the value of the interaction strength, and since this has some dependence on the motional state high fidelity requires molecules cooled to the motional ground state. While this is feasible~\cite{Liu2019, Caldwell2020b}, it is a weakness of this approach. Moreover, for molecules in separate tweezers, the strength of the dipole-dipole interaction limits gate times to about 1~ms. This could be shortened to $\sim 50\,\mathrm{\mu s}$ using a state-dependent optical tweezer \cite{Caldwell2020c, Caldwell2021} or lattice where pairs of molecules can be brought much closer, though this adds experimental complexity. 

The dipole-dipole interactions between Rydberg atoms can be far stronger than for molecules~\cite{Urban2009}, and the Rydberg blockade gate is insensitive to the value of the interaction strength. Exploiting these advantages, fast, high fidelity entanglement between atoms has been demonstrated~\cite{Levine2019,Graham2019,Madjarov2020}. In this system, however, the trapping potential is typically turned off or altered during Rydberg excitation, and this leads to heating of the atoms. This may limit the number of gate operations unless state-independent traps can be formed~\cite{Wilson2022,Zhang2020b} or non-destructive cooling techniques can be implemented~\cite{Belyansky2019}.

Here, we propose a hybrid system of molecules and atoms that exploits the advantages of each system for quantum computing. The interaction between molecules and Rydberg atoms has previously been proposed for cooling~\cite{Huber2012} and detecting molecules~\cite{Kuznetsova2016, Zeppenfeld2017,Zhelyazkova2017, Jarisch2018}. Previous work also proposed an entangling gate by placing a molecule inside the electron orbit of a Rydberg atom, producing a shift in the atomic excitation frequency that depends on the state of the molecule~\cite{Kuznetsova2011}. In this case, the Rydberg energy depends strongly on the molecule-core separation (see Fig. 2 of ref.~\cite{Kuznetsova2011}), so the method is vulnerable to motion of the particles. In our scheme, which is motivated by the Rydberg blockade gate~\cite{Saffman2010} and adiabatic gate~\cite{Petrosyan2017}, molecular qubits interact indirectly though their interactions with the large dipole moments of Rydberg atoms. An atom is placed between a pair of molecules and is coupled to a Rydberg state by a laser. The atom-molecule dipole-dipole interaction blocks the excitation if either molecule is in qubit state $\ket{1}$, but not when both are in $\ket{0}$. A $2\pi$-pulse applied to the atom implements a two-qubit gate between the molecules. Gate times of about 1~$\mu$s can be achieved with fidelities well above 99\%. The gate \change{has little sensitivity} to the motional states of the molecules, there is no heating of the qubits, and the qubits do not need to be moved. Heating of the atoms during Rydberg excitation is harmless since they do not carry any quantum information; they can easily be re-cooled or replaced. Gates can be implemented between neighbouring molecules with the help of a single atom, or between distant molecules with the help of two or three atoms. The atoms can also be used to initialize and read out the qubits, without needing to drive any molecular transitions. Thus, our scheme is suitable for fault-tolerant quantum computing using a large array of molecules. 

\section{Gate between neighbouring molecules}

\begin{figure}
	\includegraphics[width=0.5\textwidth]{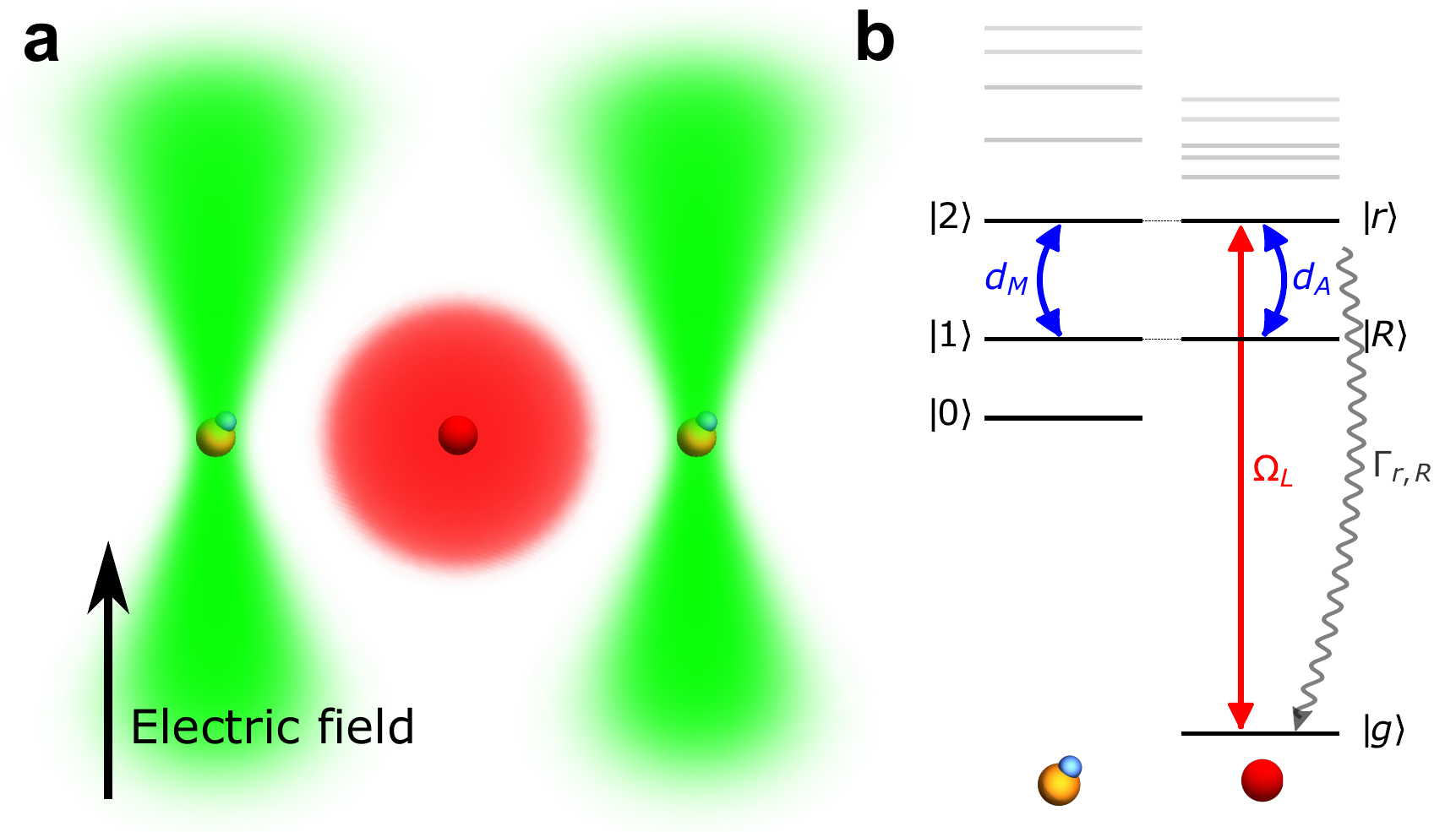}
	\caption{Proposed experimental configuration and level scheme. a. Two molecules are trapped in optical tweezers with an atom between them that can be excited to a Rydberg state. b. Level schemes (not to scale) of the molecule (left) and the atom (right). The atomic transition $\ket{r} \leftrightarrow \ket{R}$ is tuned into resonance with the molecular transition $\ket{1} \leftrightarrow \ket{2}$ by an electric field. The blue curved arrows represent transition dipole moments and the red arrow indicates the laser coupling. The faded levels indicate molecular and Rydberg states that are not needed for the gate but may result in van der Waals shifts.  }
	\label{Fig1}
\end{figure}

Figure~\ref{Fig1} illustrates our two-qubit gate scheme. Using optical tweezers, an atom is trapped between a pair of molecules. We choose an atom-molecule spacing of about $1\,\mathrm{\mu m}$. The atom trap may be switched off during the gate operation, but the molecule traps remain on. Figure \ref{Fig1}b shows the relevant levels of the molecule and the atom. Qubit states $|0\rangle$ and $|1\rangle$ are encoded in two hyperfine or rotational states of the molecule, while state $\ket{2}$ is used for the gate and is connected to $|1\rangle$ by an electric dipole transition. The atom has ground state $\ket{g}$ and two Rydberg states $\ket{r}$ and $\ket{R}$. The atomic transition $\ket{r} \leftrightarrow \ket{R}$ is tuned into resonance with the molecular transition $\ket{1} \leftrightarrow \ket{2}$ using a small electric field.  The gate is simply a smoothly-varying $2\pi$-pulse coupling $|g\rangle \leftrightarrow |r\rangle$. In the absence of any detuning, a suitable Hamiltonian describing the gate is (see Appendix \ref{sec_Hamiltonian})
\begin{align}
    &\tilde{H} = \frac{\Omega_{\rm L}}{2}\left( \ket{g}\bra{r} +\ket{r}\bra{g}\right) \nonumber \\ &+ \frac{V_{\rm dd}}{2}\left( \ket{1r}\bra{2R} + \ket{2R}\bra{1r} + \ket{1'r}\bra{2'R} + \ket{2'R}\bra{1'r}\right).
    \label{eq:H_main}
\end{align}
where $\Omega_{L}$ is the Rabi frequency, $V_{\rm dd}$ is the strength of the dipole-dipole coupling between the degenerate pair states $\ket{1r}$ and $\ket{2R}$, and we use primes to denote the states of the second molecule. We have assumed that the atom is equidistant from the two molecules, though this is not necessary. We choose $\Omega_{\rm L}(t)=\Omega_\mathrm{max}\sin{\left(\frac{\pi t}{T}\right)}$, with $t\in\left[0,T\right]$ and $\Omega_\mathrm{max}=\pi^2/T$, but any smooth function that has the same area can be used. Suitable choices of states for CaF and Rb are discussed in Appendix \ref{sec_states}, but the scheme is applicable to a wide range of molecules and atoms.

\begin{figure*}[!t]
	\includegraphics[width=0.9\textwidth]{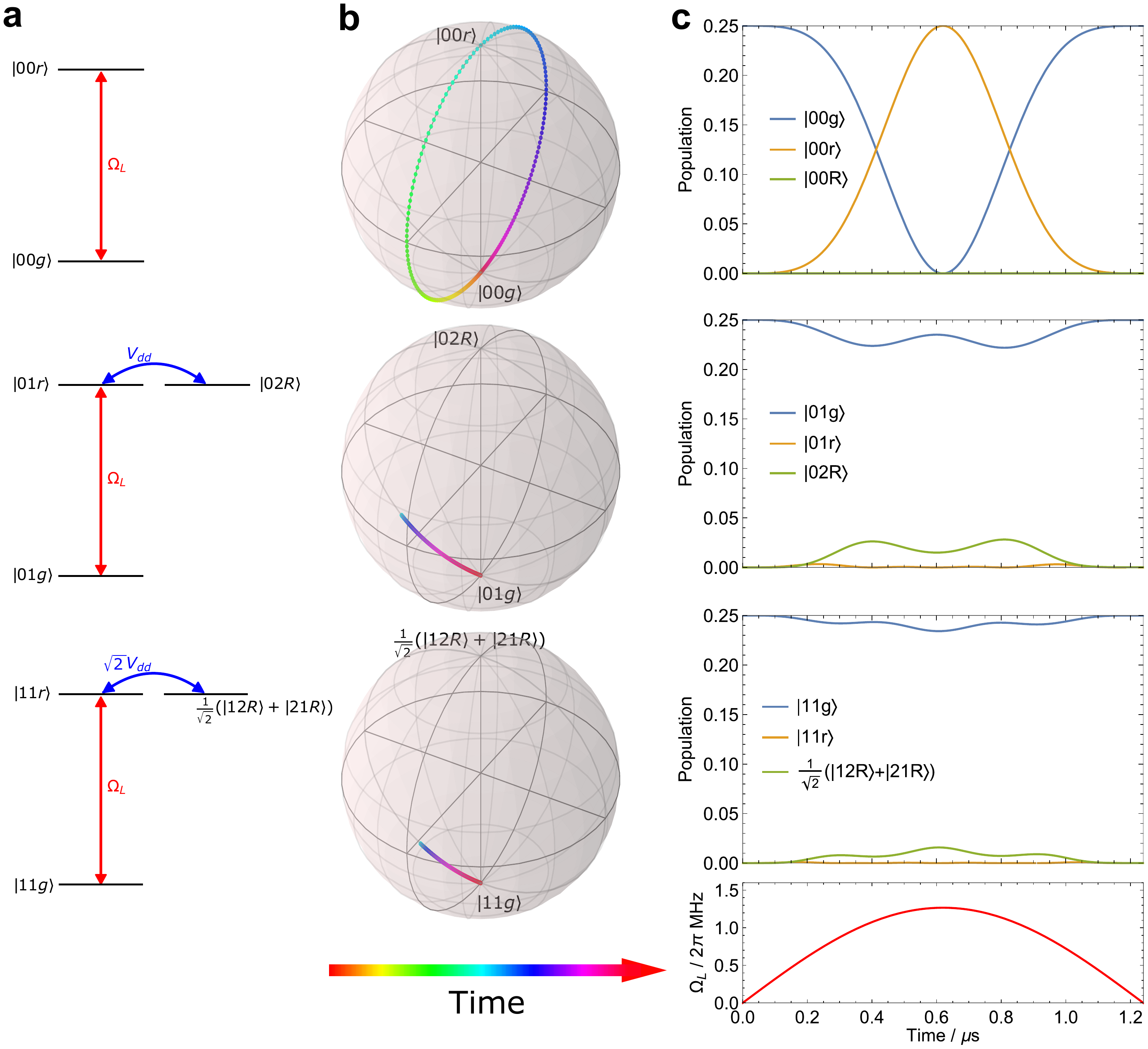}
	\caption{Gate dynamics. a. Relevant three-body states and the couplings between them. The blue arrows stand for dipole-dipole interactions. The red arrows are the laser coupling between $|g\rangle \leftrightarrow |r\rangle$. b. The evolution of the states on the Bloch sphere. Color indicates time. When the initial state is $|00g\rangle$, the system completes a Rabi oscillation, acquiring a $\pi$ phase shift. For initial states $|01g\rangle$, $|10g\rangle$ and $|11g\rangle$, the system adiabatically follows the uncoupled states that have zero amplitude for the atom in $\ket{r}$, returning along the same path on the Bloch sphere and acquiring no phase. c. The population in each state by numerical simulation. \change{The initial state is an equal superposition of all four two-qubit states.} Here, we have chosen $V_\mathrm{dd}/2=2\pi\times 2.02\,\mathrm{MHz}$, $\Omega_\mathrm{max}=2\pi\times 1.3\,\mathrm{MHz}$ and $T=1.25\,\mathrm{\mu s}$, which are suitable parameters for our example of CaF and Rb (see Appendix \ref{sec_states}) and are typical for many polar molecules.}
	\label{Fig2}
\end{figure*}

Figure \ref{Fig2} illustrates the dynamics of the gate process. When both molecules are in $\ket{0}$ (top row), the atom undergoes a complete Rabi oscillation and the system acquires a phase $\pi$, so that the three-body state evolves from $\ket{00g}$ to $-\ket{00g}$. When one of the two molecules is in $\ket{1}$ (middle row, initial state $\ket{01g}$), the system has an uncoupled eigenstate of zero energy, $\ket{u}=-\sin\theta \ket{01g}+\cos\theta\ket{02R}$, where $\tan \theta = V_{\rm dd}/\Omega_{L}$ (see Appendix \ref{sec_Hamiltonian}). When $\Omega_{L}$ changes adiabatically, which is equivalent to $\Omega_{\rm max} \ll V_{\rm dd}$, the system starts in $\ket{u}$ and remains in $\ket{u}$ throughout, returning to $\ket{01g}$ at the end of the pulse without acquiring a phase. When both molecules are in $\ket{1}$ (bottom row), the process is similar, but now the uncoupled state is formed from the states $\ket{11g}$ and $\ket{\Psi^+} = \frac{1}{\sqrt{2}}(\ket{12R}+\ket{21R}$). We see that the system acquires a phase $\pi$ only when both molecules are in $\ket{0}$. This implements the controlled-Z (CZ) gate $U_1=-|00\rangle \langle 00|+|01\rangle \langle 01|+|10\rangle \langle 10|+|11\rangle \langle 11|$, which in combination with single-qubit gates is sufficient for quantum computing.

\begin{figure*}
	\includegraphics[width=0.8\textwidth]{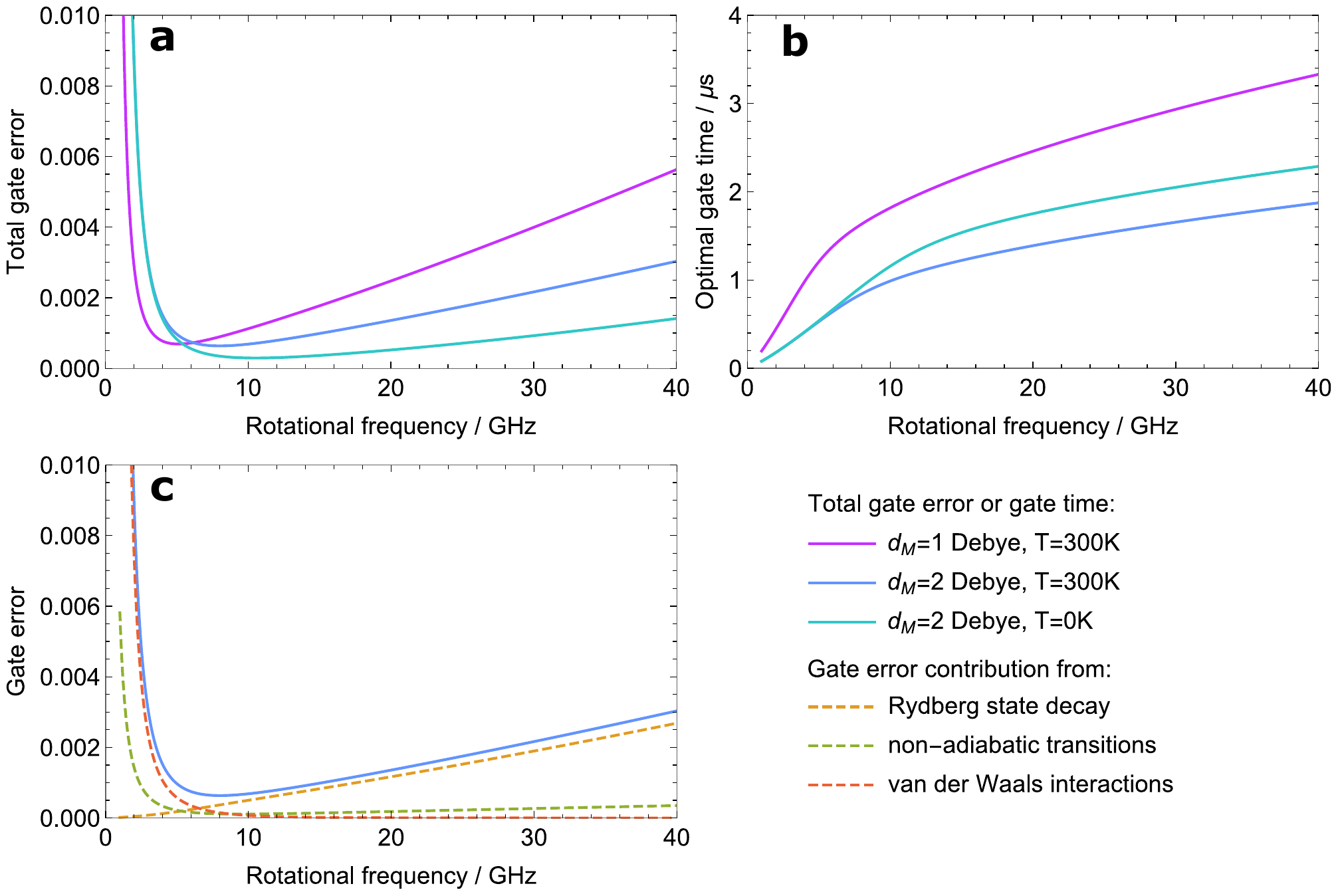}
	\caption{Gate performance as a function of the rotational frequency of the molecule, for two different molecular dipole moments ($d_{\rm M}$) and environmental temperatures, $T$. We have assumed an atom-molecule separation of $1\,\mathrm{\mu m}$ with an uncertainty of $0.2$~$\mu$m. a. Total gate error when the gate time is optimized. b. Optimum gate time. c. Error contributions from three different sources when $d_{\rm M}=2$~Debye and $T=300$~K.}
	\label{Fig3}
\end{figure*}

Next, we consider the duration and fidelity of the gate. We focus on four sources of error -- Rydberg decay, non-adiabatic processes, van der Waals interactions, and detunings. The error due to Rydberg decay is reduced for shorter gate times, whereas non-adiabatic errors are smaller for longer gate times, so there is an optimum gate duration whose value depends on the Rydberg lifetime and on $V_{\rm dd}$. Both increase with the principal quantum number $n$ of the Rydberg state, so the gate duration and gate error both reduce by using states of high $n$. However, the high density of states (faded levels in Fig.~\ref{Fig1}) and large transition dipole moments at high $n$ produce substantial van der Waals interactions between the atom and molecule, resulting in a level shift that depends sensitively on the atom-molecule separation, $x_{\rm AM}$. The tweezer trap holding the atom is turned off during the gate, leading to some uncertainty ($\delta x$) in $x_{\rm AM}$ which is approximately the size of the atom's motional wavefunction. The resulting uncertainty in the level shift gives a phase error which becomes important at high $n$.  Several imperfections can contribute to gate error by detuning the laser frequency from the $\ket{g} \leftrightarrow \ket{r}$ transition or introducing a detuning between the molecular transition ($\ket{1}\leftrightarrow \ket{2}$) and the Rydberg transition ($\ket{r}\leftrightarrow \ket{R}$). The main source of detuning is likely to be fluctuations of the applied electric field. In Appendix \ref{sec_error}, we provide detailed analytical calculations of these gate errors, as well as several more minor sources of error, and we evaluate the errors for a set of realistic parameters using \change{CaF and Rb as an example, finding a total error of $6 \times 10^{-3}$ in this case.} 

Figure \ref{Fig3} shows results of these calculations. Here, we plot the gate error and optimum duration as a function of the rotational frequency of the molecule ($f$, the frequency of $\ket{1} \leftrightarrow \ket{2}$). Since the rotational and Rydberg transitions must be resonant, higher $f$ implies lower values of $n$. We see from the figure that the optimum gate duration increases with increasing $f$. This is because higher $f$ (lower $n$) reduces $V_{\rm dd}$, requiring longer gate times to keep the non-adiabatic error small. This in turn increases the gate error due to Rydberg lifetime, which is the dominant error at higher $f$. At low $f$, the gate can be very short but the error due to van der Waals interaction dominates and becomes large. While this contribution can be reduced with better localization of the particles (smaller $\delta x$), it will always tend to dominate at sufficiently low $f$. Figure \ref{Fig3} also shows how the gate error and duration depend on the molecular dipole moment, $d_{\rm M}$, and the temperature of the environment, $T$. Increasing $d_{\rm M}$ increases $V_{\rm dd}$ which reduces both the gate time and the gate error at higher frequencies but increases the dominant van der Waals error at low frequencies. Reducing $T$ increases the Rydberg lifetimes which reduces the error for higher frequencies. We have not included the detuning error in Fig.~\ref{Fig3} since it is a technical error whose value depends on the laser stability and electric field stability achievable. For the example considered in Appendix \ref{sec_states}, the detuning error is $3\times 10^{-3}$ when the field stability is 1 part in $10^{4}$. \change{We have also neglected effects due to photon recoil during Rydberg excitation~\cite{Robicheaux2021} since these effects are typically smaller than the ones considered and can be eliminated by using a three-photon transition~\cite{Ryabtsev2016}.}

Importantly, we see from Fig.~\ref{Fig3} that over a broad range of parameters the gate error is well below $10^{-2}$ and the gate time is of order 1~$\mu$s. \change{Indeed, the figure shows that when $d_{\rm M}=2$~Debye and $T=300$~K, the error is below $10^{-3}$ for rotational frequencies between 4 and 16 GHz. These parameters coincide with the dipole moments and frequencies of low-lying rotational transitions for many molecules.} The gate is about $10^{4}$ times faster than gates using the direct dipole-dipole interaction between molecules at a similar spacing. Furthermore, since the gate relies only on adiabatic following, it is insensitive to the value of $V_{\rm dd}$ and to the exact quantum state evolution. This makes it insensitive to position fluctuations or motion of the particles, with the exception of the van der Waals error which is negligible unless $f$ is particularly small. In addition, since the energies of all relevant three-body states are independent of the particle separations during the gate, there is no force between the particles to heat up the system. Numerical simulations of the gate, described in Appendix \ref{sec_simulation}, confirm our values for the gate error, and show that the gate is indeed robust to motion and \change{introduces negligible heating. The simulations predict that when the molecules are in a thermal state with mean phonon number $\bar{n}$, the gate error arising from the motion is approximately $2 \times 10^{-4} \sqrt{\bar{n}}$. For a typical trap frequency of 100 kHz, $\bar{n}=1$ corresponds to a temperature of 5~$\mu$K. This temperature has already been reached in experiments with molecules formed by atom association~\cite{Cairncross2021}, and with directly cooled molecules~\cite{Cheuk2018, Caldwell2019, Ding2020}. Thus, error due to motion is negligible for molecules in readily achievable thermal states.}

\change{\section{Gates within arrays of molecules}}

\begin{figure*}
	\includegraphics[width=\textwidth]{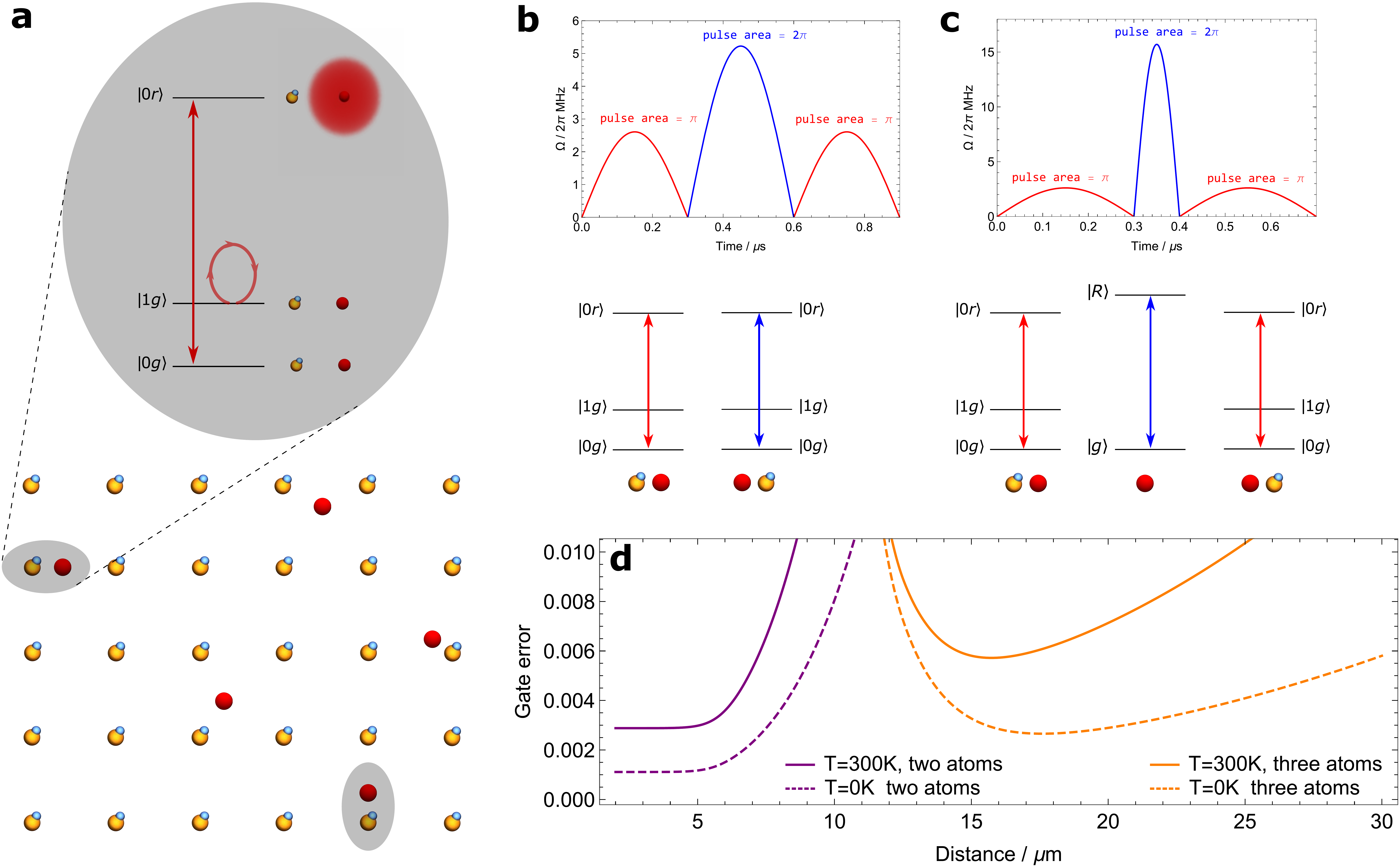}
	\caption{Long-range entangling gate. a. In a two-dimensional array of molecular qubits, atoms are moved close to the target pair of molecules (shaded areas). As before, a laser pulse will excite the atom to the Rydberg state when its molecule neighbour is in state $\ket{0}$, but not when in state $\ket{1}$.  b. Pulse sequence and level scheme for gate operation using two atoms. Red pulses are applied to atom 1, blue pulse to atom 2.  c. Pulse sequence and level scheme for gate operation using three atoms. Red pulses are applied to both outer atoms, blue pulse to central atom. d. Gate error for the two-atom and three-atom protocols. Here, we have used the example of a Rb atom located 0.6~$\mu$m from a CaF molecule, resulting in $V_{\rm dd}/2=2\pi\times 9.5$~MHz, as discussed in Appendix \ref{sec_longrange}. }
	\label{Fig4}
\end{figure*}

Our scheme becomes especially powerful when implemented within an array of molecules and extended so that operations can be done between distant qubits in the array. In a uniform 2D array, resonant interactions with next-nearest neighbours can result in an unacceptably large gate error. \change{We propose three possible solutions to this problem. The first is simply to move the chosen pair closer together for the gate. Fast coherent transport of entangled atoms in tweezer arrays has recently been demonstrated~\cite{Bluvstein2022}, and similar methods could be used for molecules. A second option, which avoids any movement of the qubits, makes use of the fact that different molecular states can have different light shifts. In the ideal case, the light shifts of $\ket{0}$ and $\ket{1}$ should be identical so that the qubit frequency is insensitive to the tweezer intensity and to the motional state, while the light shift of $\ket{2}$ should be substantially different from $\ket{1}$ so that the tweezer intensity can be used to control the detuning of $\ket{1} \leftrightarrow \ket{2}$ from $\ket{R} \leftrightarrow \ket{r}$. In this way, all molecules can be shifted off resonance apart from the targeted pair. The dipole-dipole coupling of the Rydberg atom to the next-nearest neighbour in the array is $V_{\rm NNN}/2 = V_{\rm dd}/(2\times 5^{3/2})$. When this is small compared to the detuning introduced by the change of tweezer intensity, $\Delta_{\rm NNN}$, the transition $\ket{00g} \leftrightarrow \ket{00r}$ will be detuned by $V_{\rm NNN}^2/4\Delta_{\rm NNN}$ when the state of the next nearest neighbour is $\ket{1}$, but not when it is $\ket{0}$. This introduces a phase error into the gate. In Appendix \ref{sec_states} we find suitable states for CaF molecules, and in Appendix \ref{sec_error} we find that the gate error due to next nearest neighbour interactions is $7 \times 10^{-4}$ when $\Delta_{\rm NNN}=2\pi\times 2$~MHz. For typical trap depths, a detuning of this magnitude can be achieved by approximately doubling (or halving) the intensity. A third way to avoid errors from next nearest neighbours is to introduce an extra state $\ket{1^*}$. Then, $\ket{0}$ and $\ket{1^*}$ are storage qubits whereas $\ket{1}$ and $\ket{2}$ are used for the gate as before. A mapping from $\ket{1^*}$ to $\ket{1}$ is used to prepare the selected qubits for the gate. Since all other molecules remain in $\ket{1^*}$, which is chosen to have no resonant dipole-dipole coupling with the Rydberg atom, interactions with next nearest neighbours are eliminated. In Appendix \ref{sec_states} we choose suitable states in CaF molecules to illustrate how this method can work in practice.} 

Using one of these steps, the gate can be applied to any neighbouring pairs in the array without significant errors arising from other nearby molecules. Importantly, the gate can also be extended to long distances by using two or three atoms, as illustrated in Fig.~\ref{Fig4}. We consider a pair of distant molecules, each with an atom trapped nearby (shaded areas in Fig.~\ref{Fig4}a). As before, a molecule blocks excitation of the nearby atom when in state $\ket{1}$, but not when in state $\ket{0}$. In addition, only one of the two atoms can be excited to the Rydberg state due to the van der Waals interaction between them; this is the usual Rydberg blockade. These two mechanisms work together to implement a long-range CZ gate between the molecules.  The gate protocol is illustrated in Fig.~\ref{Fig4}b: A $\pi$-pulse is applied to the first atom, then a $2\pi$-pulse to the second atom, and finally another $\pi$-pulse to the first atom. All three pulses are resonant with $\ket{g}\leftrightarrow\ket{r}$ and have smoothly evolving Rabi frequencies. If the initial state of the two qubits is $\ket{00}$ or $\ket{01}$, the first atom (alone) completes one Rabi cycle. If the state is $\ket{10}$, the second atom completes a Rabi cycle. If the state is $\ket{11}$, neither atom is excited. This implements the CZ gate $U_2=-|00\rangle \langle 00|-|01\rangle \langle 01|-|10\rangle \langle 10|+|11\rangle \langle 11|$. The additional gate error from adding an extra atom and from imperfect Rydberg blockade is small when the interaction between two Rydberg states is much stronger than the Rydberg-molecule interaction. This extends the gate to $10\,{\rm \mu m}$ for our example of CaF and Rb, and even further if higher Rydberg states are used. 

\begin{figure*}
	\includegraphics[width=\textwidth]{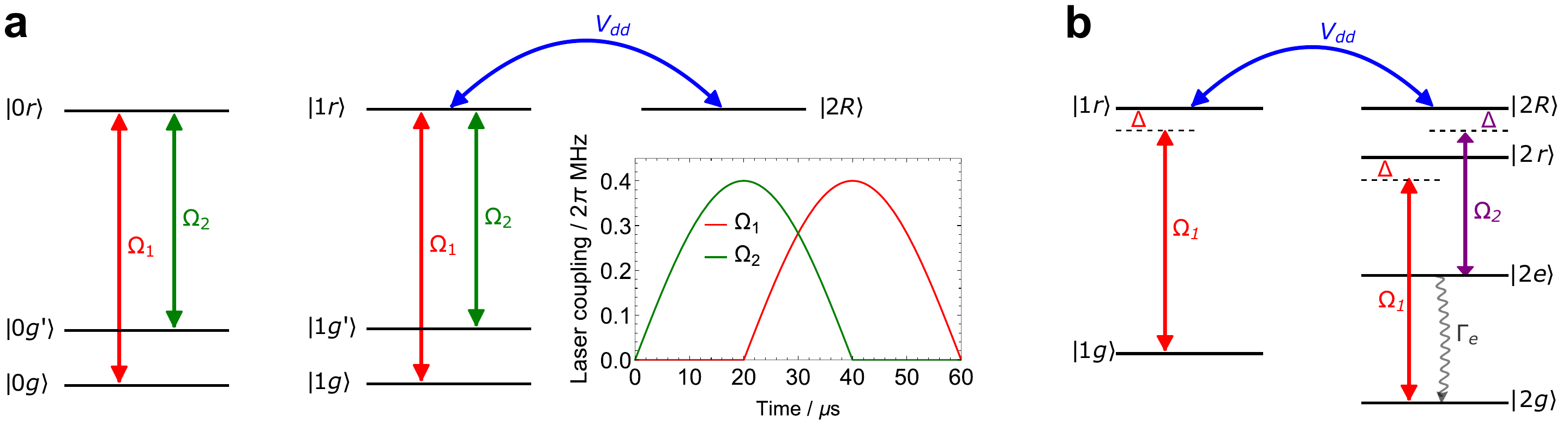}
	\caption{a. Qubit readout. The atom has two hyperfine ground states $\ket{g}$ and $\ket{g'}$. Two laser fields, $\Omega_1$ and $\Omega_2$, couple $\ket{g}\leftrightarrow \ket{r}$ and $\ket{g'}\leftrightarrow \ket{r}$, respectively. The STIRAP pulse sequence shown in the plot is applied. The atom initially in $\ket{g}$ is adiabatically transferred to $\ket{g'}$ when the molecule is in $\ket{0}$ (left), but this transfer is suppressed by the dipole-dipole interaction (blue arrow) when the molecule is in $\ket{1}$. Subsequently, the state of the atom is read out using a fluorescence measurement. b. Qubit initialization. Two laser fields $\Omega_1$ and $\Omega_2$ couple $\ket{g}\leftrightarrow \ket{r}$ and an excited state (not a Rydberg state) $\ket{e}\leftrightarrow \ket{R}$ with the same detuning. We first map $\ket{0}$ to $\ket{2}$. The state $\ket{1g}$ is resonantly coupled to $\ket{2e}$ by $\Omega_1$, $V_{\rm dd}$ and $\Omega_2$, and $\ket{2e}$ decays rapidly to $\ket{2g}$. The system cannot be excited out of $\ket{2g}$ due to the large detuning. Finally, $\ket{2}$ is mapped back to $\ket{0}$.}
	\label{Fig5}
\end{figure*}

The dipole blockade utilised above relies on the van der Waals interaction between two atoms in $\ket{r}$. We can instead make use of the resonant dipole-dipole interaction between a pair of Rydberg atoms, one in $\ket{r}$ and the other in $\ket{R}$. This has a longer range so can extend the gate to molecules at even larger separations. To do this, we add a third atom, somewhere in the middle, that connects the two atom-molecule units (Fig.~\ref{Fig4}a). The gate protocol is shown in Fig.~\ref{Fig4}c: A $\pi$-pulse resonant with $\ket{g}\leftrightarrow \ket{r}$ is applied to the two outer atoms, followed by a $2\pi$-pulse on the $\ket{g}\leftrightarrow \ket{R}$ transition of the central atom, and finally another $\pi$-pulse applied to the two outer atoms. The key point here is that the atom pair states $\ket{rR}$ and $\ket{Rr}$ are coupled by the dipole-dipole interaction, so if one of the outer atoms is in $\ket{r}$, the central atom will adiabatically follow the uncoupled eigenstate formed from $\ket{rg}$ and $\ket{Rr}$, returning to $\ket{rg}$ at the end of the $2\pi$-pulse without acquiring a phase. This is the same as the atom-molecule dynamics illustrated in Fig.~\ref{Fig2}. If the initial state of the two molecules is $\ket{00}$, both outer atoms are excited and complete a Rabi cycle. If the state is $\ket{01}$ or $\ket{10}$, only one of the outer atoms is excited, and if the state is $\ket{11}$ only the central atom is excited. In all cases, all three atoms return to the ground state at the end of the pulse sequence. The protocol implements a long-range CZ gate $U_3=|00\rangle \langle 00|-|01\rangle \langle 01|-|10\rangle \langle 10|-|11\rangle \langle 11|$.  

Figure \ref{Fig4}d shows the gate error as a function of distance for these long-range gates. For separations below 10~$\mu$m the two-atom gate has the lower error. That error is dominated by Rydberg decay at short distance, and by imperfect Rydberg blockade at longer distances. The three atom gate works well for distances between 12 and 25~$\mu$m. At shorter distances the van der Waals interaction (Rydberg blockade) inhibits the simultaneous excitation of the two outer atoms, while at longer distances it is harder to maintain the non-adiabatic condition, leading to longer gate times and increased Rydberg decay. More details about these long-range gates and their errors can be found in Appendix \ref{sec_longrange}. 

\section{Rydberg-assisted qubit readout and initialization}

We have focussed on the two-qubit gate, but the same methods can also be used to initialize and read out the molecular qubits without any direct laser interaction with the molecules. This is a major advantage, because direct methods for detecting the state of a molecule are typically destructive. Ultracold molecules formed by atom association are typically detected by dissociating them back into atoms. Laser cooled molecules can be detected by laser-induced fluorescence, but this results in heating and potentially the loss of molecules from the tweezer traps. These destructive methods can be circumvented by using the resonant dipole-dipole interaction between the molecule and a Rydberg atom.

Figure \ref{Fig5}a shows an example of how to read out the qubit state using the atom. The atom has a pair of hyperfine components $\ket{g}$ and $\ket{g'}$, which are coupled by lasers to the same Rydberg level $\ket{r}$. The atom is initially in $\ket{g}$. Stimulated Raman adiabatic passage (STIRAP) using the pulse sequence shown in the figure transfers the atom to $\ket{g'}$ when the qubit state is $\ket{0}$. We calculate a probability exceeding 0.9999 for this step. When the qubit state is $\ket{1}$ the STIRAP is blocked by $V_{\rm dd}$ and the atom remains in $\ket{g}$. Here again, the probability is greater than 0.9999. Subsequently the state of the atom is measured with high fidelity by laser-induced fluorescence detection on a cycling transition. 

The method described above can also be used to initialize the qubit state -- after the measurement all qubits that are in $\ket{1}$ are rotated to $\ket{0}$. An alternative initialization scheme that does not rely on detection is shown in Fig.~\ref{Fig5}b. Here, we introduce an additional state of the atom, $\ket{e}$, which is a state that decays rapidly to $\ket{g}$. The method is as follows. First, we transfer molecules from $\ket{0}$ to $\ket{2}$. Next, we use two laser fields (Rabi frequencies $\Omega_1$ and $\Omega_2$) to connect $\ket{g}$ to $\ket{r}$ and $\ket{e}$ to $\ket{R}$ with identical detuning $\Delta$. When the molecule is in state $\ket{1}$, $\ket{1g}$ is resonantly coupled to $\ket{2e}$ through the combination of $\Omega_1$, $V_{\rm dd}$ and $\Omega_2$. Since $\ket{e}$ decays rapidly, the system is optically pumped to $\ket{2g}$. Once the molecule reaches state $\ket{2}$ nothing further happens since $\ket{2g}$ is sufficiently far detuned from $\ket{2r}$. Finally, we map $\ket{2}$ back to $\ket{0}$ so that all molecules are now initialized in $\ket{0}$.

\section{Conclusions}

We have presented a hybrid system that combines the advantages of molecules and Rydberg atoms for quantum computing. Our two-qubit gate takes about 1~$\mu$s, which is $10^{3}$ -- $10^4$ times faster than other gate protocols for molecules. A gate fidelity as high as 99.9\% is feasible. The gate protocol is robust to motion -- the molecules can be in thermal motional states and the atoms do not need to be trapped during Rydberg excitation. The gate does not heat up the qubits and the gradual heating of the atoms \change{has a negligible effect} since they are not carrying the quantum information. The principle of the gate is simple and it can be applied to a wide range of polar molecules produced in tweezers either by atom association~\cite{Zhang2020} or direct cooling~\cite{Anderegg2019}. We have shown how to extend this gate to long-range, and find that it can work well for separations of at least 25~$\mu$m, and longer for higher Rydberg states or at cryogenic temperatures. This provides connectivity of qubits across a large array without ever needing to move them. We have shown how the atoms can be used to initialize and read out the qubits, avoiding the need to dissociate or scatter photons from the molecules, thus circumventing completely the difficulties associated with complex molecular energy level structures. Together, these techniques provide a complete set of operations for fast and scalable quantum computation using molecules.

\change{{\it Note added} -- We wish to draw attention to related work~\cite{Wang2022} which we became aware of while completing this manuscript.}

\begin{acknowledgements}
We are grateful for support from the European Commission (101018992) and from EPSRC (grants EP/P01058X/1, EP/V011499/1, EP/W00299X/1).
\end{acknowledgements}
\vspace{1cm}

Author contributions. C.Z. proposed the initial idea. Both authors did the calculations and analysis, and contributed to discussions and writing of the manuscript.

\onecolumngrid
\appendix

\section{Hamiltonian}
\label{sec_Hamiltonian}
States of the molecule are labelled $\ket{0}$, $\ket{1}$ and $\ket{2}$. States of the atom are labelled $\ket{g}$, $\ket{R}$ and $\ket{r}$. States $\ket{0}$ and $\ket{g}$ have zero energy. In cases where we need to distinguish between the states of the two molecules, we write the states of the second molecule as $\ket{0'}$, $\ket{1'}$ and $\ket{2'}$. Energies are $\omega_{i}$ with $i \in \{1,2,R,r\}$. The $\ket{1} \leftrightarrow \ket{2}$ transition is near-resonant with the $\ket{R} \leftrightarrow \ket{r}$ transition and we define $\Delta_{\rm AM} = (\omega_2-\omega_1) - (\omega_r - \omega_R)$. A laser of frequency $\omega_{\rm L}$ is near-resonant with the $\ket{g} \leftrightarrow \ket{r}$ transition with detuning $\Delta_{\rm L}=\omega_{\rm L}-\omega_{r}$ and Rabi frequency $\Omega_{\rm L}(t)$.  The atomic and molecular parts of the Hamiltonian are 
\begin{align}
    H_{\rm A} &= \omega_r \ket{r}\bra{r} + \omega_{R}\ket{R}\bra{R}, \\
    H_{\rm M} &= \omega_1 \ket{1}\bra{1} + \omega_2 \ket{2}\bra{2}.
\end{align}
The atom-light interaction is
\begin{equation}
    H_{\rm AL} = \frac{\Omega_{\rm L}}{2}\left( e^{i \omega_{\rm L} t} \ket{g}\bra{r} + e^{-i \omega_{\rm L} t} \ket{r}\bra{g} \right).
\end{equation}
The dipole-dipole interaction is 
\begin{equation}
H_{\mathrm{dd}} = \frac{1}{4\pi\epsilon_0} \left( \frac{\boldsymbol{d}_{\rm A}\cdot\boldsymbol{d}_{\rm M} - 3(\boldsymbol{d}_{\rm A}\cdot\mathbf{n})(\boldsymbol{d}_{\rm M}\cdot\mathbf{n})}{|\boldsymbol{x}_{\rm A}-\boldsymbol{x}_{\rm M}| ^3} \right),
\label{dd_interaction}
\end{equation}
where $\boldsymbol{d}_{\rm A,M}$ are the electric dipole moments for the atom (A) and molecule (M), $\boldsymbol{x}_{\rm A,M}$ are their positions, and $\mathbf{n}$ is a unit vector in the direction of $\boldsymbol{x}_{\rm A}-\boldsymbol{x}_{\rm M}$. $H_{\mathrm{dd}}$ does not change the energies of the atoms or molecules to first order, but couples different pair-states. The interaction between molecule 1 and the atom is
\begin{equation}
    H_{1 {\rm A}} = \frac{V_{1 \rm{A}}}{2}\left( \ket{1r}\bra{2R} + \ket{2R}\bra{1r} \right),
\end{equation}
where
\begin{equation}
    V_{1 \rm{A}} = 2 \bra{1r}H_{\rm dd} \ket{2R},
\end{equation}
and we have neglected the off-resonant part of the interaction [terms $\ket{1R}\bra{2r}$ and $\ket{2R}\bra{1r}$]. A similar term describes the interaction between molecule 2 and the atom, with coefficient $V_{2A}$. When the atom is exactly half way between the two molecules, $V_{1A}=V_{2A}=V_{\rm dd}$. We have used $V_{\rm dd}$ throughout the main part of the paper. We can neglect the molecule-molecule interaction which is about 4 orders of magnitude smaller than the molecule-atom interaction. This is because the molecules are twice as far apart and, for a typical choice of Rydberg state, the atomic transition dipole moment is roughly 1000 times larger than the molecular one.

We transform the Hamiltonian using the operator $U=U_{\rm M}U_{\rm M}U_{\rm A}$ where
\begin{align}
    U_{\rm M} &= \exp\left( i\omega_{1} t \ket{1}\bra{1} + i \omega_2 t \ket{2}\bra{2} \right), \\
    U_{\rm A} &= \exp\left( i\omega_{\rm L} t \ket{r}\bra{r} + i (\omega_{\rm L}+\omega_{1}-\omega_2) t \ket{R}\bra{R} \right).
\end{align}
The transformed Hamiltonian is
\begin{align}
    \tilde{H} &= -\Delta_{L}\ket{r}\bra{r} + (\Delta_{\rm AM}-\Delta_{L}) \ket{R}\bra{R} + \frac{\Omega_{\rm L}}{2}\left( \ket{g}\bra{r} + \ket{r}\bra{g}\right) \nonumber \\ &+ \frac{V_{\rm 1A}}{2}\left( \ket{1r}\bra{2R} + \ket{2R}\bra{1r}\right) +
    \frac{V_{\rm 2A}}{2}\left( \ket{1'r}\bra{2'R} + \ket{2'R}\bra{1'r}\right).
    \label{eq:H_final}
\end{align}

Consider the subsystem formed by $\ket{01'g}$, $\ket{02'R}$ and $\ket{01'r}$, and introduce the new states
\begin{align}
    \ket{c} &= \cos\theta \ket{01'g} + \sin \theta \ket{02'R},\nonumber\\
    \ket{u} &= -\sin\theta \ket{01'g} + \cos \theta \ket{02'R},\nonumber\\
    \ket{e} &= \ket{01'r},\nonumber\\
    \tan\theta &= \frac{V_{\rm 2A}}{\Omega_{\rm L}}.\nonumber
\end{align}
In the case where $\Delta_{\rm L} = \Delta_{\rm AM}$ the Hamiltonian can be expressed as
\begin{equation}
    \tilde{H} = -\Delta_{\rm L} \ket{e}\bra{e}  + \frac{\sqrt{\Omega_{\rm L}^2 + V_{\rm 2A}^2}}{2}\left( \ket{c}\bra{e} + \ket{e}\bra{c}\right).
    \label{eq:ham_uce}
\end{equation}
The state $\ket{u}$ is an eigenstate of this Hamiltonian with zero energy. If the system starts in $\ket{01'g}$, which is identical to $\ket{u}$ when $\Omega=0$, and $\Omega$ changes adiabatically, first increasing and then decreasing back to zero, the system remains in $\ket{u}$ throughout, returning to $\ket{01'g}$ without acquiring any phase.

Similarly, we can consider the subsystem formed by $\ket{11'g}$, $\ket{11'r}$ and $\ket{\Psi^+} = \frac{\left(V_{\rm 1A} \ket{21'R} + V_{\rm 2A}\ket{12'R}\right)}{\sqrt{V_{\rm 1A}^2 + V_{\rm 2A}^2}}$. As above, we introduce the new states
\begin{align}
    \ket{c} &= \cos\theta \ket{11'g} + \sin \theta \ket{\Psi^+},\nonumber\\
    \ket{u} &= -\sin\theta \ket{11'g} + \cos \theta \ket{\Psi^+},\nonumber\\
    \ket{e} &= \ket{11'r},\nonumber\\
    \tan\theta &= \frac{\sqrt{V_{\rm 1A}^2 + V_{\rm 2A}^2}}{\Omega_{\rm L}},\nonumber
\end{align}
so that, when $\Delta_{\rm L} = \Delta_{\rm AM}$, the Hamiltonian becomes
\begin{equation}
    \tilde{H} = -\Delta_{\rm L}\ket{e}\bra{e}  + \frac{\sqrt{\Omega_{\rm L}^2 + V_{\rm 1A}^2 + V_{\rm 2A}^2}}{2}\left( \ket{c}\bra{e} + \ket{e}\bra{c}\right).
\end{equation}
Once again, since $\ket{u}$ is an eigenstate, if the system starts in $\ket{u}$ it will remain in $\ket{u}$ and acquire no phase as long as $\Omega(t)$ changes adiabatically. Note that this does not require equal coupling of the atom to the two molecules, so their separations do not have to be identical.

\section{Choice of atomic and molecular states and applied electric field}
\label{sec_states}

Let us consider the properties that the qubit and auxiliary states should have. First, the qubit states $\ket{0}$ and $\ket{1}$ should ideally have identical Zeeman shifts and identical light shifts so that the qubit frequency is insensitive to fluctuations of the magnetic field and the tweezer intensity. Second, the transition dipole moment between $|0\rangle$ and $|1\rangle$ should be zero so that $\ket{00}$, $\ket{01}$, $\ket{10}$ and $\ket{11}$ are all eigenstates of the two-molecule Hamiltonian including the dipole-dipole interaction. A third requirement is a strong transition dipole moment between $|1\rangle$ and $|2\rangle$, implying that they should be in neighbouring rotational states of the molecule, and a strong transition dipole moment between $\ket{r}$ and $\ket{R}$. This will ensure a large resonant dipole-dipole interaction leading to fast and robust gates. To understand the fourth requirement, note that while the atom-molecule pair state $\ket{1r}$ can be coupled to other pair states without affecting the adiabatic following of the uncoupled state, near resonant coupling between $\ket{2R}$ and any other pair states can shift the energy of the uncoupled state or eliminate it altogether. Thus, we should ensure that $H_{\rm dd}$ cannot couple $\ket{2R}$ to $\ket{1''r''}$ where $\ket{1''}$ and $\ket{r''}$ are Zeeman sublevels degenerate with $\ket{1}$ and $\ket{r}$ respectively. To work out how to satisfy this last requirement, we need to analyze $H_{\rm dd}$ for our choice of geometry. We choose the applied electric field (which is along $z$) to be perpendicular to the plane of the 2D array. The $x$-axis can then be anywhere in the plane, and we choose it to be along $\bf{n}$, the line joining the atom and molecule. The dipole-dipole interaction, Eq.~(\ref{dd_interaction}), then has the form
\begin{equation}
    H_{\mathrm{dd}} = \frac{1}{4\pi\epsilon_0 x_{\rm AM}^3} \left( d_{\rm{A},0}d_{\rm{M},0} + \frac{1}{2}(d_{\rm{A},-1}d_{\rm{M},+1}+d_{\rm{A},+1}d_{\rm{M},-1})-\frac{3}{2}(d_{\rm{A},-1}d_{\rm{M},-1}+d_{\rm{A},+1}d_{\rm{M},+1}) \right)
    \label{eq:Hdd_cartesian}
\end{equation}
where $x_{\rm AM}=|\boldsymbol{x}_{\rm A}-\boldsymbol{x}_{\rm M}|$ and $d_{{\rm A(M)},p}$ is the spherical tensor component $p$ of the dipole operator for the atom (molecule). This form makes it clear that a $\pi$ transition on the atom couples only to a $\pi$ transition in the molecule, while a $\sigma^{\pm}$ transition of the atom will couple to both a $\sigma^{\pm}$ and a $\sigma^{\mp}$ transition of the molecule.

\change{Tensor Stark shifts are often large for molecules. They lead to decoherence due to fluctuations of the tweezer intensity or motion in the trap, so it is best to choose qubit states where these shifts are small, as noted above. Conversely, the tensor shifts can be very useful for driving single-qubit and two-qubit gates on targeted molecules within an array. Thus, it is useful for state $\ket{2}$ to have a substantial tensor Stark shift. By changing the tweezer intensity, targeted molecules can be selectively brought into resonance with a microwave field used to drive single-qubit operations, or selectively brought into resonance with a neighbouring Rydberg atom for two-qubit operations. The tensor Stark shift also lifts the degeneracy with respect to $|m_F|$ which can be useful in shifting unwanted dipole-dipole couplings out of resonance. To set the scale, we note that a typical tweezer trap with trap frequency of order 200~kHz and waist size close to the tweezer wavelength will have a trap depth of roughly 40~MHz. When non-zero, the tensor Stark shifts are typically 10-20\% of the scalar shifts. Thus, changes in transition frequencies of a few MHz can be achieved by doubling or halving the tweezer intensity. }

\begin{figure}
	\includegraphics[width=0.7\textwidth]{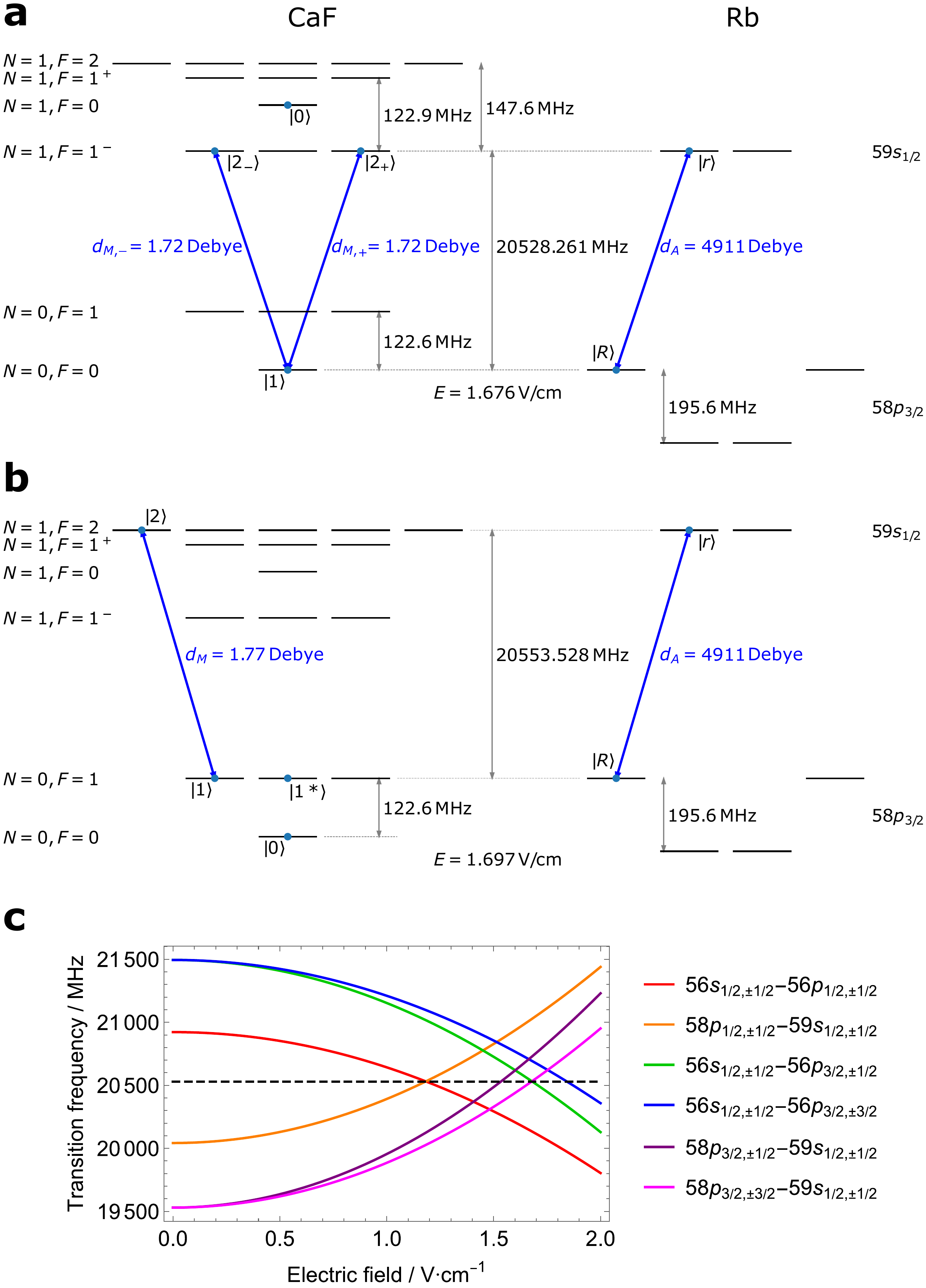}
	\caption{\change{a,b. Two possible implementations using CaF and Rb showing the relevant energy levels and choices of states. The Rydberg states $\ket{r}$ and $\ket{R}$ are the same in both implementations. In a, $\ket{0}$ and $\ket{1}$ are qubit states, while $\ket{1} \leftrightarrow \ket{2}$ is used for the gate. Individual molecules are targeted by changing the intensity of the tweezer, utilizing the tensor Stark shift of the gate transition. In b, $\ket{0}$ and $\ket{1^*}$ are qubit states, while $\ket{1} \leftrightarrow \ket{2}$ is used for the gate. Individual molecules are targeted by mapping $\ket{1^*}$ to $\ket{1}$ prior to the gate. c. Tuning the Rydberg transitions to resonance with the molecule transition. The dashed line marks the frequency of the molecular transition $\ket{1} \leftrightarrow \ket{2}$. The Stark shift of this transition is invisible on this scale. The slight difference between the molecular transition frequencies in a and b is not shown. Colored curves are Rydberg transitions which are easily tuned into resonance with the molecular transition.  $\ket{1} \leftrightarrow \ket{2}$ is resonant with $\ket{r} \leftrightarrow \ket{R}$ when the electric field is close to $1.7\,\mathrm{V/cm}$.}}
	\label{Fig6}
\end{figure}

With these criteria in mind, we now look at the specific example of a CaF molecule and a Rb atom. Figure \ref{Fig6} shows the relevant energy levels. On the left hand side, we see the hyperfine structure within the first two rotational states of the molecule. Here, $N$ and $F$ are the quantum numbers for the rotational angular momentum and total angular momentum respectively.\footnote{Within $N=1$ there are two states that have $F=1$ which we  distinguish with a superscript $\pm$.} \change{We label the states using the notation $\ket{N,F,m_F}$. The figure shows two different choices for the qubit and gate states, and we consider each in turn. In Fig.~\ref{Fig6}a we choose $\ket{0} \equiv \ket{1,0,0}$, $\ket{1} \equiv \ket{0,0,0}$ and $\ket{2_{\pm}} \equiv \ket{1,1^{-},\pm 1}$. As desired, the qubit states have no first-order Zeeman shifts, small differential light shifts, and no transition dipole moment between them. State $\ket{2}$ has a large tensor Stark shift, and this can be used to tune the next-nearest neighbour interactions out of resonance by changing the tweezer intensity for the targeted molecules.} The transition frequency from $\ket{1}$ to $\ket{2_\pm}$ is shown by the dashed line in Fig.~\ref{Fig6}c. Many Rydberg transitions of Rb are close to this frequency. Here, we focus on the $ns \leftrightarrow np$ and $np \leftrightarrow (n+1)s$ transitions because they have large transition dipole moments. The figure shows the Stark shifts of several Rydberg transitions of interest, which can all be brought into resonance with the molecule by applying an electric field between 1 and 2~V/cm. We choose\footnote{In this notation, the first subscript is $J$ and the second is $m_J$.} $\ket{r}=\ket{59s_{1/2,-1/2}}$ and $\ket{R}=\ket{58p_{3/2,-3/2}}$. Using the information about the allowed couplings given above, we see that $H_{\rm dd}$ couples $\ket{1r}$ to both $\ket{2_+R}$ and $\ket{2_-R}$, but that $\ket{2_{\pm}R}$ have no near-resonant couplings to any other pair state. Using the values of the transition dipole moments given in Fig.~\ref{Fig6}a, together with Eq.~(\ref{eq:Hdd_cartesian}), we find that when $x_{\rm AM}=1$~$\mu$m, $\frac{V_\mathrm{dd,+}}{2} = \langle 1r|H_{\mathrm{dd}} |2_+ R\rangle = -2\pi \times 0.64\,\mathrm{MHz}$ and $\frac{V_\mathrm{dd,-}}{2} = \langle 1r|H_{\mathrm{dd}} |2_- R\rangle = 2\pi \times 1.91\,\mathrm{MHz}$. Introducing $|2\rangle \equiv \frac{1}{\sqrt{V_{\rm dd,+}^2+ V_{\rm dd,-}^2}}(V_\mathrm{dd,+}|2_+\rangle + V_\mathrm{dd,-}|2_-\rangle)$, we see that $|1r\rangle$ is coupled to $\ket{2R}$ with an interaction strength of $\frac{V_\mathrm{dd}}{2} = \langle 1r|H_{\mathrm{dd}} |2R\rangle = \frac{1}{2} \sqrt{V_\mathrm{dd,+}^2+V_\mathrm{dd_-}^2} =2\pi \times 2.02\,\mathrm{MHz}$.

\change{Figure \ref{Fig6}b shows a second possible implementation. Here, we choose $\ket{0} \equiv \ket{0,0,0}$ and $\ket{1^*} \equiv \ket{0,1,0}$ as qubit states. Again, these satisfy the criteria of having no first-order Zeeman shifts, no vector or tensor light shifts, and no transition dipole moment between them. For the gate, we choose $\ket{1} \equiv \ket{0,1,-1}$ and $\ket{2} \equiv \ket{1,2,-2}$. A two-photon transition is used to map $\ket{1^*}\leftrightarrow \ket{1}$ before and after the gate. This targets the molecules of interest, eliminating interactions from other molecules in the array. We use the same Rydberg transition as before. In this case, $H_{\rm dd}$ couples $\ket{1r}$ to $\ket{2R}$ and also to $\ket{2''R}$ where $\ket{2''} \equiv \ket{1,2,0}$. This latter coupling is 7.3 times smaller and is shifted off resonance by the tensor part of the ac Stark shift. For a typical trap depth of 40~MHz, the shift is about 9~MHz, meaning that this extra coupling can be neglected. Thus, considering only the transitions shown in Fig.~\ref{Fig6}b, we calculate $\frac{V_{\rm dd}}{2} = 2\pi\times 1.96$~MHz at $x_{\rm AM}=1$~$\mu$m. The coupling strengths are almost the same in the two schemes, so in analyzing the performance of the gate we typically do not need to distinguish between the two implementations. }

In this section we have looked at a specific example which illustrates the practicality of the gate. However, we emphasize that our gate scheme will work for a wide range of molecules and Rydberg atoms.

\section{Gate error analysis}
\label{sec_error}

In this Appendix we calculate the gate error from various sources. The gate error is defined as 
\begin{equation}
    \epsilon = 1-\left( \mathrm{Tr}\sqrt{\sqrt{\rho} \rho_{\rm target} \sqrt{\rho}} \right)^2,
\end{equation}
where $\rho_{\rm target}$ is the target state and $\rho$ is the final state after the gate. The laser pulse is $\Omega_{\rm L}(t) = \Omega_\mathrm{max} \sin (\pi t/T)$, where $T$ is the gate time and $\Omega_\mathrm{max}=\frac{\pi^2}{T}$. The results are summarized in Table~\ref{table1}.

\begin{table}
\caption{Gate error and scaling with molecule properties.}
\label{table1} 
\begin{center}
\begin{tabular}{ c|c c c } 
\hline
Error source & Simplified expression\footnote{The expressions are simplified by setting $V_\mathrm{dd}T=10\pi$ to fix the adiabatic condition. See the complete expressions in text.} & Scaling\footnote{$V_\mathrm{dd}\propto d_Ad_M \propto n^{2} d_M$, level spacing $\Delta E_r \propto n^{-3}$, rotational frequency needs to match Rydberg transition frequency, so $f\propto n^{-3}$. The decay rates $\Gamma\propto n^{-5/2}$ (this is the relevant scaling at $T=300$~K for $50 \lesssim n \lesssim 80$, which corresponds to $7~\mathrm{GHz} \lesssim f \lesssim 33~\mathrm{GHz}$. At $T=0$, the scaling is close to $n^{-3}$).} & Error for CaF \\
\hline
Rydberg state decay & $(2.7\Gamma_r+0.9\Gamma_R)/V_\mathrm{dd}$ & $f^{3/2} d_M^{-1}$ & $1.4\times10^{-3}$ \\ 
Non-adiabatic transitions & $9\pi^6/(16V_\mathrm{dd}^4T^4)$ & - & $5.5\times10^{-4}$ \\ 
van der Waals interaction\footnote{We assume $\delta x/x_{\rm AM} = 0.2$.} & $0.52V^{2}_\mathrm{dd}/(E_{\rm ryd}^2)$ & $f^{-10/3}d_M^2$ & $1\times 10^{-6}$ \\ 
Detuning\footnote{Assuming electric field fluctuations dominate and the fractional error in the electric field is fixed.} & $22.4\Delta^2/V_\mathrm{dd}^2$ & $f^{10/3} d_M^{-2}$ & $3.0\times 10^{-3}$ \\
\change{Next nearest neighbours}\footnote{\change{This error applies when the tensor Stark shift is used to detune the next-nearest neighbours. We assume $\Delta_{\rm NNN} = 2\pi\times 2$~MHz.}} & $1.8 \times 10^{-4} V_{\rm dd}^2/\Delta_{\rm NNN}^2$ & $f^{-4/3} d_M^2$ & $7.2\times 10^{-4}$\\
Molecule-molecule interaction & - & $d_M^2$ & $< 10^{-10}$\\
External field induced interaction & - & $f^{-2/3} d_M^2$ & $< 10^{-9}$ \\ 
Total error & - & - & $5.7\times10^{-3}$ \\
\hline
\end{tabular}
\end{center}
\end{table}

\subsection{Rydberg state decay}
\label{sec:Ryd_decay}

The gate error caused by Rydberg state decay is $\sum_{r'} \Gamma_{r'} \int p_{r'}(t) dt $ where the sum is over all relevant Rydberg states $r'$, $p_{r'}(t)$ is the probability of the system being in $r'$ at time $t$, and $\Gamma_{r'}$ is the decay rate of $r'$.

The initial state $|00g\rangle$ is resonantly coupled to $|00r\rangle$ by the laser, and the population in $|r\rangle$ (or $|00r\rangle$) is $p_{00r}(t)=\sin^2\left(\int_{0}^{t} \frac{\Omega_{\rm L}(t')}{2} \,dt'\right)$. The integrated population is $P_{00r}=\int_0^T p_{00r}(t)\,dt = \frac{1}{2}\left( 1+J_0(\pi)\right)T$, where $J_{n}(x)$ is the Bessel function. When starting in $|01g\rangle$ or $|10g\rangle$, the state follows the dark eigenstate $ V_\mathrm{dd} |01g\rangle - \Omega_{\rm L}(t) |02R\rangle$ and the population in the Rydberg state is $p_{02R}(t)=\frac{\Omega_{\rm L}(t)^2}{V_\mathrm{dd}^2+\Omega_{\rm L}(t)^2}$. The integrated population is $P_{02R}= \zeta(V_{\rm{dd}}T)T$ where
\begin{equation}
    \zeta(x) = 1-\frac{1}{\sqrt{1+\left(\frac{\pi^2}{x}\right)^2}}.
\end{equation}
A similar expression can be written  for the case where the initial state is $|11g\rangle$. After accounting for the probabilities of being in each of the initial states, the total decay error is
\begin{equation}
\begin{split}
\epsilon_\Gamma &= \frac{1}{4}\left(P_{00r}\Gamma_r + 2P_{02R}\Gamma_R+P_{12R}\Gamma_R\right)\\
&=\frac{1}{8}\left( 1+J_0(\pi)\right)\Gamma_r T + \frac{1}{4}\left(2\zeta(V_{\rm dd}T)+\zeta(\sqrt{2}V_{\rm dd}T)\right)\Gamma_{R}T.
\end{split}
\end{equation}

In our example of CaF and Rb, the Rydberg state lifetimes are $\tau_r=97\,\mathrm{\mu s}$ and $\tau_R=126\,{\rm \mu s}$ when the environment temperature is 293~K. For a gate time $T=1.25\,{\rm \mu s}$, this error is around $1.4\times 10^{-3}$.

\subsection{Break-down of adiabatic condition}

In our gate protocol, the initial states $|01g\rangle$, $|10g\rangle$ and $|11g\rangle$ should adiabatically follow the uncoupled eigenstates, returning to the intial state without acquiring a phase. Here, we calculate the error due to non-adiabatic transitions. We assume there is no laser detuning ($\Delta_{\rm L} = 0$); a detuning causes a different gate error which is discussed in Sec.~\ref{sec_detuning}.

We consider the case where the initial state is $\ket{01g}$. The dynamics can be described in the basis of $|u\rangle$, $|c\rangle$ and $|e\rangle$ which are formed from $|01g\rangle$, $|01r\rangle$ and $|02R\rangle$ as described in Appendix~\ref{sec_Hamiltonian}. The Hamiltonian is Eq.~(\ref{eq:ham_uce}) (with $\Delta_{\rm L}=0$), whose eigenstates are $|u\rangle$ and $|\pm\rangle=\frac{1}{\sqrt{2}}(|c\rangle\pm|e\rangle)$ with energies $0,\pm\frac{1}{2}\sqrt{V_\mathrm{dd}^2+\Omega_{\rm L}(t)^2}$. Since the Hamiltonian is time-dependent, so too are the eigenstates. The rate of change of $\ket{u}$ is $\frac{d}{dt}|u\rangle=(-\cos{\theta}|01g\rangle-\sin{\theta}|02R\rangle) \frac{d\theta}{dt}=-|c\rangle \frac{d\theta}{dt}$, where  $\tan{\theta}=V_\mathrm{dd}/\Omega_{\rm L}(t)$. This change couples $|u\rangle$ to $|\pm\rangle$ with a coupling strength $C_\pm = \langle\pm|\frac{d}{dt}|u\rangle = -\frac{1}{\sqrt{2}}\frac{d\theta}{dt} = \frac{1}{\sqrt{2}} \frac{V_\mathrm{dd}}{V_\mathrm{dd}^2+\Omega_{\rm L}(t)^2} \frac{\pi^3}{T^2}\cos\left({\frac{\pi t}{T}}\right)$. The resulting loss of population from $|u\rangle$ is $P_{\mathrm{NA},01g}=|\int_0^T C_\pm(t) \exp [i\int_0^t E_\pm(\tau) \,d\tau]\,dt|^2$ where $E_\pm$ are the energy differences between $\ket{u}$ and the $\ket{\pm}$ eigenstates~\cite{Messiah2014}. This can be approximated to $P_{\mathrm{NA},01g}\approx \frac{C_\pm^2}{4E_\pm^2}=\frac{\pi^6}{V_\mathrm{dd}^4 T^4}$ when $V_\mathrm{dd} \gg \Omega_{\rm L}(t)$. 
When the initial state is $|11g\rangle$, the population loss is a factor of 4 smaller because the interaction strength is  $\sqrt{2}V_\mathrm{dd}$. Taking into account the probabilities of the initial states, we find that the total non-adiabatic error is
\begin{equation}
\epsilon_\mathrm{NA} = \frac{9}{16}\frac{\pi^6}{V_\mathrm{dd}^4T^4}.
\end{equation}
Using the parameters from Sec.~\ref{sec_states} and a gate time $T=1.25\,{\rm \mu s}$, this error is $5.5\times 10^{-4}$.

\subsection{Detuning}
\label{sec_detuning}

The ideal gate has the laser on resonance with $|g\rangle\leftrightarrow|r\rangle$, and the pair states $|1r\rangle$ and $|2R\rangle$ are tuned to be degenerate. Unwanted detunings arise from laser frequency fluctuation or imperfect control of the Rydberg energies ($\omega_{r,R}$). We use the definitions in Appendix~\ref{sec_Hamiltonian}: the energies of $|g\rangle$, $|0\rangle$ and $|1\rangle$ are zero, $\Delta_{\rm AM} = (\omega_2-\omega_1) - (\omega_r - \omega_R)$, and $\Delta_{\rm L}=\omega_{\rm L}-\omega_{r}$.

When the initial state is $|00g\rangle$, a laser detuning shifts the energy of $|00r\rangle$ by $-\Delta_{\rm L}$, giving rise to an extra phase of $\delta\phi_{00}=-\Delta_{\rm L}P_{00r}$, where $P_{00r}=\frac{1}{2}\left( 1+J_0(\pi)\right)T$ is the integrated probability given in Sec.~\ref{sec:Ryd_decay} . This phase, when it is small, results in a gate error of $\frac{3}{16}\delta\phi_{00}^2$. Similarly, when the initial state is $|01\rangle$ or $\ket{10}$, the system passes through the state $\ket{02R}$ and acquires a phase $\delta\phi_{01,10} = (\Delta_{\rm AM}-\Delta_{\rm L})P_{02R}$. The phase is the same for the two states leading to a gate error of $\frac{1}{4}\delta\phi_{01,10}^2$. A similar result is obtained when the initial state is $\ket{11}$. Adding together the gate errors due to detunings, we obtain
\begin{equation}
\begin{split}
\epsilon_\Delta &= \frac{3}{16} \Delta_{\rm L}^2 P_{00r}^2 + (\Delta_{\rm AM}-\Delta_{\rm L})^2 \left( \frac{1}{4}P_{02R}^2 + \frac{3}{16}P_{12R}^2 \right)\\
&=\frac{3}{64}\left( 1+J_0(\pi)\right)^2\Delta_{\rm L}^2T^2 + \left(
\frac{1}{4} \zeta(V_{\rm dd}T)^2 + \frac{3}{16} \zeta(\sqrt{2}V_{\rm dd}T)^2 \right)(\Delta_{\rm AM}-\Delta_{\rm L})^2 T^2.
\end{split}
\end{equation}
In reality the phase changes are correlated, which reduces the total error, so the above expression can be taken as an upper limit.

The largest contribution to the detuning is likely to come from changes, $\Delta E$, in the applied electric field, $E$. The change in laser detuning with electric field comes from the change in the energy of $\ket{r}$ with electric field, which for our choice of states and operating field is $-457 \Delta E/E$~MHz. Similarly, the change in $(\Delta_{\rm AM}-\Delta_{\rm L})$ with electric field comes from the change in the energy of $\ket{R}$ with electric field, $-2400 \Delta E/E$~MHz. If the electric field is stable to 1 part in $10^{4}$, the resulting gate error will be $\epsilon_\Delta = 3 \times 10^{-3}$. Better stability, and thus smaller errors, could be achieved by using higher Rydberg states as electric field sensors and actively correcting the applied field.

\subsection{van der Waals interaction}
\label{sec_interaction}

As discussed in Sec.~\ref{sec_detuning}, detunings contribute to the gate error. The energy of an atom-molecule pair state such as $|0r\rangle$ or $|2R\rangle$ can be shifted by off-resonant couplings with other pair-states; this is the van der Waals interaction, which is the second order effect of $H_{\rm dd}$. The energy shifts due to the van der Waals interaction can be compensated by tuning the laser frequency and electric field so that the detunings at the equilibrium positions are tuned to zero. However, the interaction is strongly dependent on the separation of the particles ($V_{\rm vdW}\propto x_{\rm AM}^{-6}$). The tweezer trap for the atom is turned off for the gate, resulting in a position uncertainty $\delta x$ which is approximately the size of the motional wavefunction prior to turning off the trap. Consequently, the interaction shifts the energies differently at different positions leading to a dephasing error.

For our choice of states in CaF and Rb, we have estimated the van der Waals shifts for all the relevant pair states. For each pair state of interest, $i$, we calculated the energy defects $\Delta E_{ij}$ and the dipole-dipole coupling $V_{ij}/2$ for all other atom-molecule pair states ($j$) with $\Delta E_{ij} < 5$~GHz. The van der Waals shift of $i$ is then $V_{{\rm vdW}, i}=\sum_j V_{ij}^2/(4\Delta E_{ij})$. For $x_{\rm AM} = 1$~$\mu$m, we calculate $V_{\rm vdW, 0r}/h \approx -1.1$~kHz, $V_{\rm vdW, 0R}/h \approx 31$~kHz, $V_{\rm vdW, 1R}/h \approx 1.9$~kHz, $V_{\rm vdW, 2R}/h \approx 0.9$~kHz. The gate error due to these small shifts is negligible.

However, for molecules with lower rotational frequencies (corresponding to higher Rydberg states), the van der Waals interaction can be much larger due to the smaller level spacing and larger transition dipole moments. We will find an approximate expression that shows how the error caused by this interaction scales with the rotational frequency of the molecule. Here, we neglect hyperfine structure since its size can be radically different for different molecules, and because it does not scale with rotational frequency.

The variation of $V_{\rm vdW}$ arising from the spread of positions $\delta x$ is $\delta V_{\rm vdW} = 6 V_{\rm vdW}  \delta x/x_{\rm AM}$. The interaction tends to be dominated by pair states with couplings similar to the $\ket{1r} \leftrightarrow \ket{2R}$ coupling, $V_{ij} \approx V_{\rm dd}$, which scales as $n^2$. The energy defect is typically the difference between the $(n-1)p \leftrightarrow n s$ and the $n s \leftrightarrow n p$ transitions of the Rydberg atom, which we denote as $E_{\rm ryd}$. It is approximately 3~GHz at $n=59$ and it scales roughly as $n^{-3}$. With these approximations, we have $\delta V_{\rm vdW} \approx 3/2 (V_{\rm dd}^2/E_{\rm ryd})(\delta x/x_{\rm AM})$. If we assume that $V_{\rm dd} \gg \Omega_{\rm max}$, the integrated probabilities are dominated by $P_{00r}$ and we obtain an approximate gate error of
\begin{equation}
    \epsilon_{\rm vdW} \approx \frac{27 V_{\rm dd}^4}{256 E_{\rm ryd}^2} \left(\frac{\delta x}{x_{\rm AM}}\right)^2 \left( 1+J_0(\pi)\right)^2 T^2.
\end{equation}
The gate time is inversely proportional to $V_{\rm dd}$. As an example, let us take $V_{\rm dd} T = 10\pi$, giving
\begin{equation}
    \epsilon_{\rm vdW} \approx 13 \left(\frac{V_{\rm dd}}{E_{\rm ryd}}\right)^2 \left(\frac{\delta x}{x_{\rm AM}}\right)^2.
\end{equation}
Finally, we note that $V_{\rm dd}$ scales as $E_{\rm ryd}^{-2/3}$ resulting in $\epsilon_{\rm vdW} \propto E_{\rm ryd}^{-10/3}$. Using this scaling, and taking a typical value of $\delta x/x_{\rm AM} =0.2$, we estimate that the gate error will exceed $10^{-3}$ when $E_{\rm ryd} < 0.4$~GHz, corresponding to rotational frequencies below about 3~GHz.

\change{The van der Waals interaction can also be viewed as a state-dependent heating of the atom's motion, leading to a reduction in the overlap between the initial and final motional states of the atom, which then reduces the coherence in the reduced density matrix of the molecule pair. Using the values of the van der Waals shifts given above, together with the integrated populations in the Rydberg states, we estimate that this reduction is about $4\times 10^{-6}$ when the two-qubit state is $\ket{10}$ or $\ket{01}$, and much smaller for the other two states. This is a different way of viewing the same error already considered above.}

\change{
\subsection{Interactions with other molecules in the array}

We consider a square array of molecules separated by 2~$\mu$m, with an atom located half way between one pair such that $x_{\rm AM}=1$~$\mu$m. The interaction of the atom with its two nearest neighbours is $V_{\rm dd}/2 = 2\pi\times 2$~MHz. The coupling strength of the atom with one of its next-nearest neighbours (NNN) is $V_{\rm NNN}/2 = V_{\rm dd}/(2\times 5^{3/2})$. The effect of this unwanted coupling is reduced by changing the tweezer intensity for the target molecules relative to all other molecules in the array, so that only the target molecules are resonant with the atom. Then, the dipole-dipole interaction with the NNN is detuned by $\Delta_{\rm NNN}$. This uses the differential light shift of states $\ket{1}$ and $\ket{2}$ (see the discussion in Appendix \ref{sec_states}). The main influence of this off-resonant coupling occurs when the qubit state is $\ket{00}$, so that the atom undergoes a Rabi oscillation. In the limit where $\Delta_{\rm NNN} \gg V_{\rm NNN}$, the effect can be modelled as a detuning of the Rabi oscillation by $\delta_{\rm NNN} = V_{\rm NNN}^2/4 \Delta_{\rm NNN}$. This results in a detuning error of exactly the same form as discussed in Sec.~\ref{sec_detuning}. After accounting for the fact that there are four NNNs, but the error only occurs when a NNN is in state $\ket{1}$, we obtain a gate error of
\begin{equation}
    \epsilon_{\rm NNN} = \frac{3}{32}(1+J_0(\pi))^2 \delta_{\rm NNN}^2 T^2.
\end{equation}
When the qubit state is not $\ket{00}$, the error arising from the other molecules in the array is negligible compared to this error. A numerical model that includes the influence of the NNNs confirms this simple model when $\Delta_{\rm NNN} \ge 2\pi\times 2$~MHz. Taking $\Delta_{\rm NNN} = 2\pi\times 2$~MHz, and other parameters the same as previously, we find $\epsilon_{\rm NNN} = 7.2 \times 10^{-4}$.

The alternative way to remove the influence of other molecules in the array is to separate the gate states from the qubit states. This strategy does not require a change to the tweezer intensity when driving two-qubit gates. An example of this is illustrated in Fig.~\ref{Fig6}b. In that example, the interaction with the NNNs is not completely eliminated because $\ket{1}$ and $\ket{1^*}$ are degenerate and $\ket{1^*}$ can couple to $\ket{1,2,\pm 1}$ through the dipole-dipole interaction. For a trap depth of 40~MHz, this coupling is detuned by about 8~MHz and the gate error is much smaller than the one calculated above.

}

\subsection{Molecule-molecule interaction}

The dipole-dipole interaction between two molecules with $d_M\approx 2\,\mathrm{Debye}$ separated by $2\,\mathrm{\mu m}$ is about 75~Hz. Couplings and level shifts of such small size contribute a negligible gate error. 

\subsection{External field induced interaction}

An external electric field, $E$, is applied to shift the Rydberg transition into resonance with the molecular transition. This field polarizes both the atom and molecule, leading to a state-dependent, field-induced dipolar interaction.

Let $\alpha_r$ and $\alpha_1$ be the polarizabilities of states $\ket{r}$ and $\ket{1}$ respectively. The Stark shift of $\ket{r}$ is $\Delta W_r = \frac{1}{2}\alpha_r E^2$, and the induced dipole moment is $\mu_r = \alpha_r E = \sqrt{2 \alpha_r \Delta W_r}.$ Similarly, the induced dipole moment of the molecule is $\mu_1 = \alpha_1 E = \sqrt{2 \alpha_1^2 \Delta W_r/\alpha_r}$. The resulting dipole-dipole interaction between an atom in $\ket{r}$ and a molecule in $\ket{1}$ is $V_{r1} \approx \frac{\mu_r \mu_1}{4\pi\epsilon_0x_{\rm AM}^3} = \frac{\alpha_1 \Delta W_r}{2\pi\epsilon_0x_{\rm AM}^3}$. The result is similar for other atom-molecule states. The polarizability of the rotational ground state of the molecule is $\mu_0^2/(3B)$ where $\mu_0$ is the dipole moment in the frame of the molecule and $B$ is the rotational constant. Other rotational states have a similar (somewhat smaller) polarizability. Thus we obtain $V_{r1} \approx \frac{\mu_0^2 }{6\pi\epsilon_0x_{\rm AM}^3}\frac{\Delta W_r}{B}$. The Stark shift of either Rydberg state is similar in magnitude to the Stark shift of the $\ket{r} \leftrightarrow \ket{R}$ transition frequency, and this is a small fraction of the molecule's rotational frequency. In our example of CaF and Rb, the shift of the transition frequency is roughly $B/10$. Thus, we take $\Delta W_r/B \approx 1/10$ to be typical and obtain $V_{r1} \approx \frac{\mu_0^2 }{60\pi\epsilon_0 x_{\rm AM}^3}$. For $\mu_0 = 3$~Debye and $x_{\rm AM} = 1$~$\mu$m, this is roughly 100~Hz. Such a small interaction energy contributes negligibly to the gate error. This result is largely independent of rotational frequency and Rydberg principal quantum number so we expect it to hold for any reasonable choice of molecule and Rydberg atom.

\section{Numerical simulation including motional degrees of freedom}
\label{sec_simulation}

We have simulated the gate numerically to demonstrate that it \change{has little sensitivity} to motion and to confirm our error analysis. The model includes two molecules, each with three levels $\{|0\rangle,|1\rangle,|2\rangle\}$, and an atom with three levels $\{|g\rangle,|r\rangle,|R\rangle\}$. The initial state is $\frac{1}{4} (|00\rangle+|01\rangle+|10\rangle+|11\rangle)|g\rangle$. The Hamiltonian for the gate is Eq.~(\ref{eq:H_final}) with $\Omega_{\rm L}(t)=\Omega_\mathrm{max}\sin{\frac{\pi t}{T}}$. The distances between the atom and the two molecules are $x_1$ and $x_2$, which can be different. We do not include the decay of the Rydberg state in the simulation.

We first set $x_1=x_2=1\,\mathrm{\mu m}$ and use the parameters $V_\mathrm{dd}/2=2\pi\times 2.02\,\mathrm{MHz}$, $\Omega_\mathrm{max}=2\pi\times 1.3\,\mathrm{MHz}$ and $T=1.25\,\mathrm{\mu s}$. After the gate sequence, we see that the final state is $\frac{1}{4} (-|00\rangle+|01\rangle+|10\rangle+|11\rangle)|g\rangle$ with an error of $5\times 10^{-4}$. This is the non-adiabatic error. Then, we change $x_{1,2}$ by $\pm 0.1\,\mathrm{\mu m}$, or make them time-dependent, e.g. $x_1= (1\pm 0.1\sin{\omega t})\,\mathrm{\mu m}$, where the frequency $\omega<200\,\mathrm{kHz}$. The additional error caused by motion in this simulation is always less than $10^{-4}$.

Next, to describe the motional states more accurately, we assume that the molecules are trapped in harmonic potentials with a trapping frequency $\omega_{\rm m}$, whereas the atom is not trapped. We include three collective phonon modes in the $x$ direction (along the line between the molecules). Mode $a$ is the center of mass mode of the three particles, mode $b$ is a stretch mode in which the two molecules oscillate in phase with each other but out of phase with the atom, and mode $c$ is another stretch mode where the two molecules oscillate out of phase and the atom does not move. All three modes, $a$, $b$ and $c$, have the same trapping frequency. A superposition of modes $a$ and $b$ has zero trapping frequency and describes the motion of the untrapped atom. The deviation of the molecule-atom distances from their equilibrium values, $x_{1,2}=1\,\mathrm{\mu m}$, can be expressed as $\delta x_1 = \delta x_b+\delta x_c$ and $\delta x_2 = \delta x_b-\delta x_c$, where $\delta x_{b} = a_0\frac{b+b^\dagger}{2}$, $\delta x_{c} = a_0\frac{c+c^\dagger}{2}$, $a_0$ is the harmonic oscillator length and $b,b^\dagger,c,c^\dagger$ are the creation and annihilation operators of the $b,c$ phonon modes. The dipole-dipole interaction between molecule $i$ and the atom is $V_{iA}(\delta x_i) = \frac{C_3}{(x_i+\delta x_i)^3} \approx  \frac{C3}{x_i^3} - \frac{3C3}{x_i^4}\delta x_i + \frac{6C3}{x_i^5}\delta x_i^2$, where we have used a series expansion around $\delta x_i=0$. This result can then be expressed in terms of the phonon operators. The total Hamiltonian is $\tilde{H} + H_{\rm phonon}$, where $\tilde{H}$ is Eq.~(\ref{eq:H_final}) and $H_{\rm phonon}=\sum_{d\in\{a,b,c\}}(d^\dagger d+\frac{1}{2})$. We solve the time-dependent Schr\"odinger equation numerically for a thermal distribution of motional states with mean $\bar{n}$ and $\omega_{\rm m}=50\,\mathrm{kHz}$ or $100\,\mathrm{kHz}$. We find that the additional gate error arising from the distribution of motional states scales approximately as $\sqrt{\bar{n}}$ and is $2 \times 10^{-4}$ at $\bar{n}=1$. When $\omega_{\rm m}=100\,\mathrm{kHz}$, $\bar{n}=1$ corresponds to a temperature of 5~$\mu$K. We can expect molecules formed from atoms or directly laser cooled to be at or below this temperature, so we conclude that the error due to motion is negligible for molecules in readily achievable thermal states. We also see in this simulation that the gate does not result in any detectable heating -- the change in $\bar{n}$ is below $10^{-3}$.

\section{Long-range gate using two or three atoms}
\label{sec_longrange}

\subsection{Interactions between Rydberg atoms}

\begin{figure}
	\includegraphics[width=0.6\textwidth]{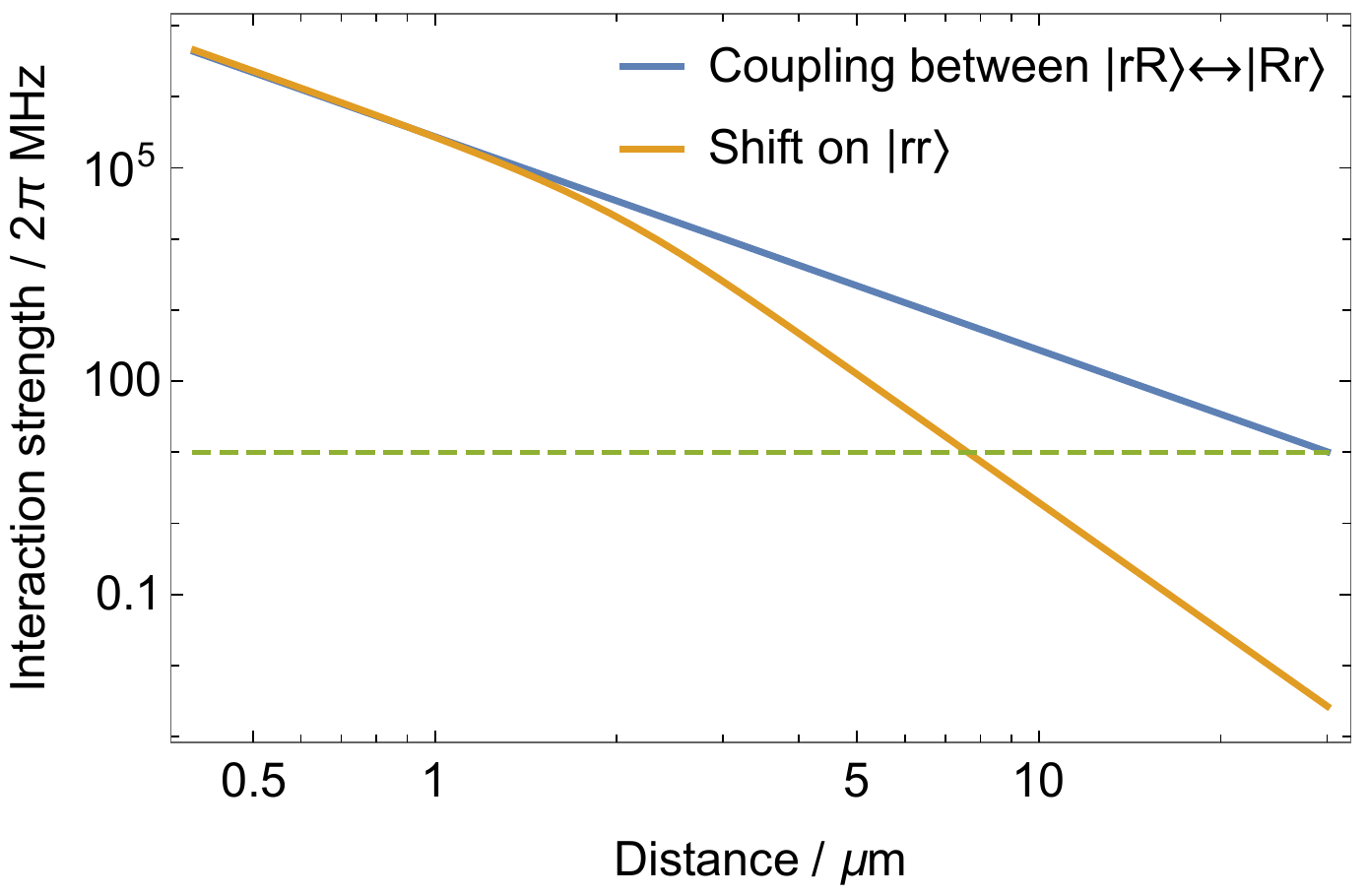}
	\caption{Interactions between atoms as a function of the distance between them. The green dashed line marks the interaction between the atom and the molecule at a separation of $0.6\,\mathrm{\mu m}$. The Rydberg states are the ones used in our example of CaF interacting with Rb.}
	\label{Fig7}
\end{figure}

In the main text, we explain how the two-qubit gate can be extended to long range by using two or three atoms. The gate using two atoms uses the van der Waals interaction between atoms in $|r\rangle$ to block the double excitation to $|rr\rangle$. This works well for a separation smaller than $10\,\mathrm{\mu m}$. For the gate using three atoms, the resonant dipole-dipole interaction between $|rR\rangle$ and $|Rr\rangle$ is used to suppress excitation of the middle atom to $|R\rangle$ when at least one of the edge atoms is in $|r\rangle$. It works for large separations where the dipole-dipole interaction $|rR\rangle \leftrightarrow |Rr\rangle$ is much stronger than the van der Waals interaction shift of $|rr\rangle$ (this energy shift causes gate error in the three-atom case). Figure \ref{Fig7} compares the strengths of these two interactions.

\subsection{Gate error}
\label{sec_longrangeerror}
Here we focus on the decay error and non-adibaticity error, and the new error sources that arise when using two or three atoms. We fix the molecule rotational frequency to 20.5\,GHz (the value for CaF) and investigate the error as a function of distance between qubits. The van der Waals error is negligible for this rotational frequency. It becomes important for frequencies below about 3\,GHz.

For two atoms, the gate consists of a sinusoidal $\pi$-pulse, $\Omega_1 = \frac{\pi^2}{2T_1} \sin (\frac{\pi t}{T_1})$, on the first atom, a sinusoidal $2\pi$-pulse, $\Omega_2 = \frac{\pi^2}{T_2} \sin (\frac{\pi t}{T_2})$, on the second atom, then a second identical sinusoidal $\pi$-pulse on the first atom. The error sources include the following. 
\begin{enumerate}
    \item Rydberg decay error. The integrated population of the first atom in $|r\rangle$ is $\frac{T_1+T_2}{2}$ ($1/2$ is the probability of starting in $|0g\rangle$ and $T_1+T_2$ is the total time in $|r\rangle$). The integrated population of the first atom in $|R\rangle$ during the adiabatic process is similar to the one-atom case,  $\frac{1}{2}\zeta(2V_{\rm dd}T_1) T_1$. Similarly, the integrated population of the second atom in $|r\rangle$ and $|R\rangle$ are $\frac{1}{8}\left( 1+J_0(\pi)\right)T_2$ and $\frac{1}{4}\zeta(V_{\rm dd}T_2) T_2$, respectively. The total decay error is 
    \begin{equation}
            \epsilon_{\Gamma 2} = \left(\frac{T_1+T_2}{2} + \frac{1}{8}\left( 1+J_0(\pi)\right)T_2\right)\Gamma_r 
            + \left(\frac{1}{2}\zeta(2V_{\rm dd}T_1) T_1 + \frac{1}{4}\zeta(V_{\rm dd}T_2) T_2\right) \Gamma_R.
    \end{equation}
    
    \item Non-adiabatic error. 
    
    The total non-adiabatic error due to the two $\pi$ pulses applied to the first atom and the $2\pi$ pulse applied to the second atom is 
    \begin{equation*}
        \epsilon_{\rm NA 2} = \frac{\pi^6}{4V_\mathrm{dd}^4T_1^4} + \frac{\pi^6}{2V_\mathrm{dd}^4T_2^4}.
    \end{equation*}
    
    \item Imperfect blockade. Both atoms may be excited to $|r\rangle$ if the energy shift $V_{rr}$ on $|rr\rangle$ is not much larger than the Rabi frequency of the excitation laser $\frac{\pi^2}{T_2}$. The resulting gate error is 
    \begin{equation*}
        \epsilon_{\rm B 2} = \frac{\pi^4}{2V_{rr}^2T_2^2}.
    \end{equation*}
\end{enumerate}

We find that if we retain an atom-molecule separation of $x_{\rm AM}=1$~$\mu$m the error is higher than for the short-range gate and approaches $10^{-2}$, mostly because there are now two atoms to decay and they spend a longer time in Rydberg states. It is preferable to reduce $x_{\rm AM}$ a little. Since there is no longer a constraint that the atom should be between two molecules, and since the atom and molecule are in independent tweezers, closer spacings can be achieved. Here, we take $x_{\rm AM} = 0.6$~$\mu$m (see Sec.~\ref{sec_separation} for further discussion).  This is much larger than the Rydberg wavefunction so the dominant interaction is still the dipole-dipole interaction. For our example of CaF and Rb, this leads to an interaction strength of $V_{\rm dd}/2 = 2\pi\times 9.5\,\mathrm{MHz}$. Using $T_1=T_2=0.3\,\mathrm{\mu s}$, the total gate error as a function of distance between the two atoms is plotted in Fig.~\ref{Fig4} of the main text. The error is below $10^{-2}$ for distances up to about $10\,\mathrm{\mu m}$.

For longer distances, the interaction (which falls off as $x^{-6}$) becomes too weak, and the gate error becomes large because of imperfect Rydberg blockade. In this regime where both atoms can be excited, we can place a  third atom half way between the other two, and excite it to $|R\rangle$. This three-atom gate consists of a sinusoidal $\pi$-pulse, $\frac{\pi^2}{2T_1} \sin (\frac{\pi t}{T_1})$, resonant with $|g\rangle \leftrightarrow |r\rangle$ and applied simultaneously to both outer atoms, followed by a sinusoidal $2\pi$-pulse, $\frac{\pi^2}{T_2} \sin (\frac{\pi t}{T_2})$, applied to the middle atom and coupling $|g\rangle \leftrightarrow |R\rangle$, then finally another identical $\pi$-pulse applied to the two outer atoms. The error sources include the following.

\begin{enumerate}
    \item Rydberg state decay. Similar to previous calculations this is
    \begin{equation}
        \epsilon_{\Gamma 3} = \left(T_1+T_2 + \frac{1}{2}\zeta(V_{\rm rR}T_2)T_2 + \frac{1}{4}\zeta(\sqrt{2}V_{\rm rR}T_2)T_2 \right) \Gamma_r 
        +  \left(\zeta(2V_{\rm dd}T_1) T_1 + \frac{1}{8}\left( 1+J_0(\pi)\right)T_2\right) \Gamma_R,
    \end{equation}
    where $V_{\rm rR}$ is the dipole-dipole interaction between $|rR\rangle \leftrightarrow |Rr\rangle$.
    
    \item Non-adiabatic error. Each of the two outer atoms sees a $\pi$-pulse, and the error is $\frac{\pi^6}{2V_\mathrm{dd}^4T_1^4}$. For the center atom the situation is identical to the one-atom case, but the interaction is replaced by $V_{\rm rR}$. The total non-adiabatic error is 
    \begin{equation*}
        \epsilon_{\rm NA 3} = \frac{\pi^6}{2V_\mathrm{dd}^4T_1^4} + \frac{9\pi^6}{16V_\mathrm{rR}^4T_2^4}.
    \end{equation*}
    
    \item Interaction between two outer atoms. When the distance is too small, the interaction between the two outer atoms detunes the $|rr\rangle$ state, which is the intermediate state of the two atoms when the molecules are in $|00\rangle$. This results in an error of:
    \begin{equation*}
        \epsilon_{\rm I 3} = \frac{3}{16}\left( T_1+T_2\right)^2 V_{rr}^2,
    \end{equation*}
    where $V_{rr}$ is the energy shift due to the van der Waals interaction. 
\end{enumerate}
Using $T_1=0.3\,\mathrm{\mu s}$ and $T_2=0.1\,\mathrm{\mu s}$, the total gate error as a function of the distance between the two outer atoms is plotted in Fig.~\ref{Fig4} of the main text. The error is below $10^{-2}$ for distances up to $25\,\mathrm{\mu m}$ at 300~K, and up to $40\,\mathrm{\mu m}$ at 0~K.

\subsection{Controlling the separation between the molecule and the atom}
\label{sec_separation}
If the atom tweezer produces a negligible light shift for the molecule, and the molecule tweezer a negligible light shift for the atom, they can be trapped independently and thus be separated by arbitrarily small distances. This may be achieved for some atoms and molecules if the zero-polarizability wavelength of one particle can trap the other particle. This is not possible for our example of CaF and Rb, but such an ideal situation is not needed. For example, one can make a molecule trap that is repulsive for the atom (e.g. 700~nm for our example) and an atom trap that is only weakly attractive for the molecule (e.g. 850~nm). The particles can then be brought to small separations without the traps merging. A more sophisticated and general approach is to use two-color optical tweezers. For example, consider a trap formed from 734\,nm and 850\,nm beams of equal intensity. This trap is attractive for CaF, but has little effect on Rb since its polarizability is roughly equal and opposite at these two wavelengths. Similarly, 554\,nm and 850\,nm have equal and opposite polarizabilities for CaF, but their combination can trap Rb whose polarizability is much larger at 850\,nm. This brief discussion illustrates that there are many possibilities for reducing $x_{\rm AM}$ to 0.6~$\mu$m (as assumed in our analysis of the long-range gate), or even closer.

\newpage




\bibliography{references}

\end{document}